%% file: main.tex
\documentclass[twocolumn]{aastex631}
\usepackage{xcolor}
\usepackage{comment}

\newcommand{\AndersenT}{B. Andersen et al. (in preparation)}

\newcommand{\ModilimT}{O. Modilim et al. (in preparation)}

\newcommand{\ShinT}{K. Shin et al. (in preparation)}

\newcommand{\SandT}{K. Sand et al. (in preparation)}
\newcommand{\SandP}{(K. Sand et al. in preparation)}
\newcommand{\DidehbaniT}{H. Didehbani et al. (in preparation)}

\newcommand{\CurtinP}{(A. Curtin et al. in preparation)}

\newcommand{\NgP}{(C. Ng et al. in preparation)}

\begin{document}

\title{Updating the first CHIME/FRB catalog of fast radio bursts with baseband data
}
\shorttitle{CHIME/FRB Catalog 1 -- Baseband}
\input{auth.tex}
\collaboration{1000}{The CHIME/FRB Collaboration}
\shortauthors{The CHIME/FRB Collaboration}
\correspondingauthor{Daniele Michilli}
\email{danielemichilli@gmail.com}

\begin{abstract}
In 2021, a catalog of 536 fast radio bursts (FRBs) detected with the Canadian Hydrogen Intensity Mapping Experiment (CHIME) radio telescope was released by the CHIME/FRB Collaboration.
This large collection of bursts, observed with a single instrument and uniform selection effects, has advanced our understanding of the FRB population.
Here we update the results for 140 of these FRBs for which channelized raw voltage (`baseband') data are available.
With the voltages measured by the telescope's antennas, it is possible to maximize the telescope sensitivity in any direction within the primary beam, an operation called `beamforming'.
This allows us to increase the signal-to-noise ratio (S/N) of the bursts and to localize them to sub-arcminute precision.
The improved localization is also used to correct the beam response of the instrument and to measure fluxes and fluences with a $\sim 10\%$ uncertainty.
Additionally, the time resolution is increased by three orders of magnitude relative to that in the first CHIME/FRB catalog, and, applying coherent dedispersion, burst morphologies can be studied in detail.
Polarization information is also available for the full sample of 140 FRBs, providing an unprecedented dataset to study the polarization properties of the population.
We release the baseband data beamformed to the most probable position of each FRB.
These data are analyzed in detail in a series of accompanying papers. 
\end{abstract}

\section{Introduction} \label{sec:intro}
Despite a tremendous improvement in fast radio burst \citep[FRB;][]{lbm+07} searching in recent years \citep{phl21}, the progenitors of FRB sources are still debated.
We now know that one FRB-like signal has been emitted by a Galactic magnetar \citep{abb+20,brb+20} but many models remain viable to explain the rest of the population \citep{pww+18}.
For over a decade, the study of FRBs has been hindered by the small number of sources known. 
As much information as possible must be collected from a large sample of FRBs to further our knowledge about the FRB population.

CHIME/FRB, an experiment running on the Canadian Hydrogen Intensity Mapping Experiment (CHIME) radio telescope to find and study a large number of FRBs, released its first catalog containing 536 FRBs in 2021 \citep[][\emph{Catalog~1} for the remainder of the manuscript]{aab+21}.\footnote{\url{https://www.chime-frb.ca/catalog}}
This is by far the largest sample of FRB sources collected by a single instrument.
After carefully accounting for its uniform selection effects, it enabled a number of population studies to be performed \citep[e.g.,][]{jcc+21,rsl+21,ckr+22,cr22,smb+23}.
The results of Catalog~1 are based on total intensity data downsampled to a time resolution of $\sim 0.983$ ms (referred to as \emph{intensity} data).
However, for some of the events, CHIME/FRB stores the voltages measured by each one of the 1024 dual-polarization antennas of the telescope.

Voltage data -- here referred to as \emph{baseband} data -- 
measured by the receivers of a radio telescope, contain information about both the intensity and phase of the incoming radio waves \citep{lk04}.
With baseband data recorded from every receiver, the sensitivity of radio interferometers can be maximized in any direction within the primary beam of the telescope -- an operation known as \emph{beamforming} -- by applying appropriate time delays to the phase of the incoming radiation.
Beamforming can also be used to improve the localization of a source beyond the size of formed beams \citep{msn+19}, for example, by mapping the signal strength around an initial guessed position.
Finally, the frequency-dependent delay experienced by radio waves propagating through the plasma along the line of sight, an effect known as dispersion, can be coherently removed by applying the appropriate time delays as frequency-dependent phase rotations of the radio waves \citep{hr75}.
Dispersion is quantified by the dispersion measure (DM), the integrated column density of free electrons along the line of sight.

The amount of information contained in baseband data makes their processing computationally expensive.
Therefore, large field-of-view (FoV) surveys typically reduce the data rate before searching for FRBs, e.g., by converting baseband data into total intensity and downsampling the data in time.
However, many of the new-generation telescopes are able to store baseband data around the time of interesting events.
Off-line pipelines are then usually used to process the baseband data at a later stage \citep[e.g.,][]{mmm+21,scd+23}.
This is also the case for CHIME/FRB \citep{abb+18}, where for each frequency channel, $\sim 100$ ms of baseband data are saved to disk when a candidate FRB is identified.

In this work, we present the baseband data collected by CHIME/FRB for 140 FRBs between 2018 December 9 and 2019 July 1.
The discovery of these events and their analysis using downsampled intensity data have been presented in Catalog~1.
Here, we report updated results and present a release of baseband data for these 140 FRBs.
A number of follow-up studies, described in \textsection\ref{sec:further_studies}, will use baseband data to infer properties of the FRB population.

The manuscript is organized as follows.
In \textsection\ref{sec:obs}, we describe the observational setup and the processing of the data.
In \textsection\ref{sec:results}, we summarize the results obtained and give an overview of follow-up papers analyzing the FRB properties.
The full-resolution beamformed baseband data are made publicly available for additional analyses.
We describe the data release and provide a guide for accessing the data in \textsection\ref{sec:data_release}.

\section{Observations and data analysis} \label{sec:obs}
As detailed by \citet{abb+22}, CHIME is a radio interferometer consisting of four paraboloid half-cylinders with 1024 dual-polarization antennas on their focal lines.
The voltages produced by the antennas are digitized and then a
polyphase filterbank \citep{abb+22} is used to channelize the digital voltages into 1024 frequency channels in a field-programmable gate array (FPGA)-based engine \citep[the so-called \emph{F-engine};][]{bcd+16}.
The output of the F-engine is the baseband data used in the present analysis.
These baseband data are recorded for the two linear polarizations of the telescope into 1024 channels of $\sim 391$ kHz each between 400 and 800 MHz, with a time resolution of $2.56$ $\mu$s, and are stored as 4+4-bit complex integer numbers.

While baseband data are stored on a ring memory buffer for $\sim 20$ s, a real-time search for FRBs is run on the data after converting them to total intensity, as detailed in the following.
To perform the real-time FRB search, 1024 beams are formed via a fast Fourier transform in a GPU-based X-engine \citep[FFT beams;][]{nvp+17,msn+19}. 
After beamforming, and before searching for FRBs, the data are up-channelized to 16,384 channels of $\sim 24.4$ kHz each to reduce the effect of intrachannel signal smearing due to dispersion delay, then transformed to total intensity and downsampled to $\sim 0.983$ ms, as described by \citet{nvp+17} and further detailed by \citep{abb+18}.
This constitutes the so-called intensity data defined in \textsection\ref{sec:intro} and reported in Catalog~1, which is used to perform the real-time search for FRBs.

If this automated real-time pipeline identifies a potentially interesting candidate in the intensity data, from the 20-s ring buffer, $\sim 100$ ms of baseband data is dumped to disk around the signal of interest for each of the 1024 frequency channels, following the dispersion delay as measured by the real-time pipeline  \citep{abb+18}.
The exact length of the dump depends on the uncertainties on DM and time of arrival measured by the real-time pipeline.
Baseband data is then stored on disk for each of the 2048 receivers of the telescope.

Unfortunately, not all FRBs discovered by CHIME/FRB have baseband data, for different reasons.
First, the system to trigger and store baseband data started its operations on 2018 December 9, i.e., more than four months after the start of Catalog~1.
Second, to avoid storing too many false positives, a threshold on the S/N has been set to trigger a baseband dump. 
The threshold has been changed over time but it has been usually set between $\text{S/N}>10$--$12$, while the threshold to store intensity data is $\text{S/N}>8$.
Finally, in the first months of operations of the baseband system, the uncertainties on DM and time of arrival were not properly accounted for in some of the events, causing the loss of part or all of the baseband data for these bursts.
After all these selections, there are 140 FRBs from Catalog~1 with usable baseband data, 12 of which come from 7 FRB sources observed to repeat.
The events are processed uniformly, in other words, no prior information is used on the repeating bursts.
In particular, any independent knowledge of, e.g., source localization from intervening follow-ups with an interferometer goes unused in our present analysis.

\subsection{Automated baseband pipeline}
\label{sec:obs:pipeline}
An off-line automated pipeline was developed to process a large number of baseband events in a rapid and consistent way.
A detailed description is given by \citet{mmm+21} and \citet{mmm+21b}; here, we report a summary of this baseband pipeline.
The pipeline is divided into three stages: \emph{refinement}, where the initial guess position from the real-time pipeline is refined with a large grid of beams in the sky; \emph{localization}, where the position is further refined with a compact grid of beams; and \emph{single-beam}, where a single beam is formed to the best source position found in the previous step and the baseband data are stored for this sky location.
The single-beam file containing beamformed baseband data is then used to run additional studies.

The beamforming operation in each stage of the pipeline starts with a phase and amplitude calibration of the baseband data using a daily observation of a persistent source.
Each frequency channel is loaded into memory and converted to 32+32-bit complex floating-point numbers for calculations.
The required number of beams is then formed independently for each frequency channel.
Finally, all channels are merged together and stored on disk.
Analysis pipelines are then run on the beamformed baseband data stored in these files.
Initially, signal smearing due to dispersion delay is coherently removed by using the initial DM guess from the real-time pipeline \citep{hr75}.
Channels corrupted by radio-frequency interference are then masked out, and S/N is calculated for each channel by normalizing the off-pulse region.
Finally, the DM of the burst is incoherently refined by maximizing the S/N of the signal.
In the refinement and localization stages, the data are downsampled in time and only total intensity data are stored on disk, while both baseband and total intensity data are stored in the single-beam stage.

The baseband processing pipeline for Catalog~1 events is run on the computer cluster of the Canadian Advanced Network for Astronomy Research (CANFAR)\footnote{\url{https://www.canfar.net}} in a series of Docker containers.
Different containers are created for the different parts of the pipeline.
For the beamforming step, the most computationally expensive, the frequency channels are distributed among 64 containers, with 2 cores each, and the result is merged together after all are completed.

\subsection{Localization}\label{sec:analysis:localization}
The localization of FRBs by CHIME/FRB with baseband data is detailed by \citet{mmm+21}.
In summary, in the refinement stage, 53 beams covering $10\times0.2$\degr\ of the sky are formed in the frame horizontal to the telescope. 
Then, a Markov chain Monte Carlo algorithm is run to fit the S/N measured in each channel of every beam to a model of the expected telescope response at those locations. 
The free parameters of the fit are the source position and its spectrum, with the latter being modeled as a Gaussian function.
Due to the shape of the reflector, CHIME's sensitivity spreads out in the East-West direction to many degrees in its sidelobes. 
To make sure localizations reported here are performed in the right lobe, the grid of beams formed during the refinement stage covers a total of five consecutive lobes of the formed beams at the bottom of the band.
A diagnostic plot is produced by dividing the bandwidth into two halves and showing the total S/N value for each half in every beam \citep{mbb+23}. 
The diagnostic plot for each event in the catalog has been visually inspected to ensure that the reported localization is in the right lobe of the formed beams.
Events where bandwidth and S/N were not sufficient to confirm the lobe with confidence have been noted.

In the localization stage, 25 beams are formed in a square grid of $0.32\times0.32$\degr\ on the sky.
A least-squares fit is performed with a two-dimensional Gaussian function that has been empirically verified to be a good model for the response of a formed beam near its peak sensitivity.
The partial correlation of the signal in different beams is taken into account following \citet{msn+19}.
Other than random uncertainties, a number of systematic effects affect the localization.
\citet{mmm+21} have modeled some of them, such as variations in cable lengths and in the spacing between antennas due to changes in ambient temperature.
Other effects, for example, due to imperfections in the metal structure of the telescope, are difficult to model and they constitute a lower limit on the localization precision.
\citet{mmm+21} used a sample of pulsars and a repeating FRB to correct for the collective effect of the unmodeled systematics on CHIME localizations. 
They found that the best precision achievable is $\sim 11\arcsec$ in both RA and Dec.
The effect of systematics that they measured has been included in all FRB positions and their uncertainties.

\subsection{DM}
\label{sec:obs:dm}
The DM measured in the automated pipeline by maximizing the S/N of the signal is refined with \texttt{DM\_phase} \citep{smp19},\footnote{\url{https://github.com/danielemichilli/DM_phase}} an algorithm designed to dedisperse bursts with complex morphology by maximizing the signal coherence across the band.
Effectively, this yields the DM that maximizes the structure in the signal. 

The $0.983$-ms resolution of intensity data limits the DM precision that can be measured and hides shorter-duration structures in the bursts.
On the other hand, the higher time resolution of baseband data significantly improves DM values that can be measured when the S/N is large enough \citep{sbm+23}.
Although the time or frequency resolution of baseband data could be further increased (in theory up to the Nyquist frequency of the 400 MHz bandwidth receiver), this is postponed to future studies, whereas in this catalog we use channelized baseband data with $2.56$-$\mu$s time resolution and $\sim 391$-kHz frequency resolution.

\subsection{Flux and fluence}
\label{sec:obs:flux}
Flux density and fluence values reported in Catalog~1 not only have large uncertainties but they should also be interpreted as lower limits \citep{aab+21}.
This is mainly due to the low precision of the FRB positions in Catalog~1, which prevented us from correcting for the response of both the formed and primary telescope beams \citep{apb+23}.

With baseband data, it is possible to form, in software, a beam in the direction of the FRB localized in \textsection\ref{sec:analysis:localization}.
Since the localization uncertainties are typically more than one order of magnitude smaller than the size of a formed beam \citep[$\sim 0.2\degr$;][]{abb+18}, this effectively corrects for the off-axis response of formed beams.
To measure flux density and fluence with baseband data, we use the total intensity data stored in the single-beam files created in the last stage of the automated pipeline and shared in our data release. 
The amplitude of the data is calibrated with a daily observation of a steady source, as described in \textsection\ref{sec:obs:pipeline}.
Initially, we downsample the time resolution of the data to some appropriate value measured by the real-time pipeline.
Then, we identify an off-pulse region where no peak in the signal is above three times the root mean square.
The average flux density of this off-pulse region is subtracted from the whole time series, effectively removing flux density variations between the calibration and the burst detection due to differences e.g.\ in system temperature and sky brightness.
The peak flux density is then calculated by measuring the flux density value at the maximum of the pulse profile, while the fluence is obtained by integrating the pulse profile in time.

To obtain the final values of fluence and flux density, we correct for the response of the primary beam at the FRB position.
A primary beam model has been derived as part of the effort to detect baryon acoustic oscillations in intensity maps of 21-cm emission with CHIME \citep{abc+23}.
This model is obtained by deconvolving the response of the primary beam with a large number of steady point sources observed by other telescopes \citep{abc+23}.
The primary beam model covers CHIME's whole frequency range in the main lobe, which spans $\sim 2\degr$ at 800 MHz in the East-West direction.
Therefore, we exclude from this analysis sources that are localized outside of the primary beam.
The effect of the primary beam is corrected by dividing the total flux density by the average beam response at the location of the FRB.
Since the flux densities measured by the two antenna polarizations are largely independent, we calculate flux density and fluence for each polarization independently and sum the two together only at the end.

The main sources of uncertainties in flux density and fluence values come from the noise in the data and the primary beam model.
The noise is estimated as the standard deviation of the off-pulse region for each frequency channel.
Uncertainties on the beam model have been evaluated with observations of 14 bright calibrator sources whose flux densities were measured with the Very Large Array by \citet{pb17}.
The total uncertainty on the primary beam integrated over the full bandwidth is estimated to be $\sim 10\%$ of the measurement.

\subsection{Exposure}
\label{sec:obs:exposure}
CHIME/FRB's sky exposure for detected FRBs has been reported in Catalog~1. 
As detailed by \citet{chimefrbcatalog1}, the uncertainties on the exposures for each source were dominated by the large uncertainties on the source positions.
Given the improvement enabled by baseband data, we have re-evaluated the exposure estimates, taking into account updated baseband localizations.
The methodology employed in our exposure calculation is detailed by \citet{chimefrbcatalog1}. 
In summary, during telescope operation, we store various metrics indicative of the uptime and sensitivity of the CHIME/FRB system within the Catalog~1 timeframe. These metrics are subsequently integrated with a model of CHIME/FRB beams,\footnote{\url{https://chime-frb-open-data.github.io/beam-model/}} allowing us to generate healpix exposure maps, which are then queried at the coordinates of the source.
It is important to note, however, that we have not adjusted our sensitivity thresholds to accommodate these improved localizations.
Bursts located at declinations $\gtrsim$ 70\degr are detected during both the upper and lower transits of the telescope, resulting in exposure values for each transit. 
We note that, with this definition of exposure, the uncertainties only depend on the uncertainty of the source position within the exposure map. For localization regions spanning multiple pixels, we report the mean exposure value and the standard deviation as its uncertainty.
Therefore, a localization that is sufficiently precise will result in a null uncertainty on the exposure if it lies within a pixel of the healpix map.

\section{Results and discussion} \label{sec:results}
As described in \textsection\ref{sec:intro}, intensity data were released for 536 FRBs in Catalog~1 \citep{chimefrbcatalog1}.
In this work, we update the results for 140 of these FRBs with the use of baseband data and the methods described in \textsection\ref{sec:obs}.
A summary of the results is reported in Table~\ref{tab:results}, which can also be downloaded in machine-readable format from the online version of the manuscript.
These results are analyzed, discussed, and reported in greater detail in a series of studies that will be highlighted in \textsection\ref{sec:further_studies}.

\subsection{Characteristics of the FRB population}
The waterfall plots -- i.e., the signal intensity as a function of frequency and time -- for the FRB sample are shown in Fig.~\ref{fig:wfalls}.
Depending on S/N in the frequency-averaged time series, the FRBs are displayed at 81.9, 163.8, or 655.4\,$\mu$s time resolution (downsampled from 2.56 $\mu$s by factors of 32, 64, or 256, respectively), chosen to visually highlight the structure of the bursts.
As noted in previous works \citep[e.g.,][]{phl21}, the diversity in the morphological properties of the sample is remarkable.
The brightness varies greatly from burst to burst, and so do their widths, their scattering timescales, and the number of components in the profile.
The higher resolution of baseband data makes it possible to distinguish the richness of sub-structures that comprise each burst, especially when the S/N is high \citep{fmm+23}.
The higher resolution also makes visible a scattering tail present in most of the bursts, which will be quantified by \SandT.

The possibility of correcting for the response of formed beams gives us a large sample of FRB bandwidths, greatly increasing the current number of FRBs with a measured spectrum \citep{msb+19}.
Many of the bursts are broadband, being detected across all the available channels of the $400$-MHz band of CHIME.
This is in contrast to some previous studies that were limited by an unknown beam response \citep[e.g.,][]{chimefrbcatalog1}, or to bursts from repeating sources \citep[e.g.,][]{lab+17,hss+19,abb+23}, which have a different bandwidth distribution \citep[e.g.,][]{pgk+21}.
This is investigated in a paper by \SandT.
Other than being interesting to study the characteristics of the FRB population, the large emission band for many bursts has implications for the expected detection rates of future radio telescopes with ultra-wideband receivers, such as the Canadian Hydrogen Observatory and Radio-transient Detector \citep[CHORD;][observing between 300-1500 MHz]{vlg+19} and the Deep Synoptic Array 2000 \citep[DSA-2000;][observing between 700-2000 MHz]{hrw+19}, given that the S/N of detected bursts increases as the square root of the burst's bandwidth \citep{cm03}.

\subsubsection{Noticeable outliers in the sample}
While the burst morphologies will be explored in greater detail elsewhere (\SandP), there are a few outliers that are immediately noticeable.
Most bursts are formed by one or two components but a few show a more complex morphology than the rest of the sample. 
This is the case, for example, of FRBs 20181215B, 20190122C, 20190124F, 20190411C, 20190502C, 20190617A, and 20190624B.
Also, while most FRBs show either one component or multiple components under the same envelope, a few are formed by multiple bursts clearly separated in time, such as FRBs 20190411C, 20190501B, and 20190609C.
FRB 20190630D shows an upward drift on its substructures, rarely seen among the FRB population \citep[e.g.,][]{hss+19,abb+23}.
Finally, sub-structures forming FRB 20190425A are analyzed in detail, together with other bright bursts, in an upcoming paper by \citet{fmm+23}.

\subsection{Comparison with Catalog 1}
A comparison with the results of Catalog~1 for the sample of 140 FRBs is presented in the following subsections.

\begin{figure*}
\plotone{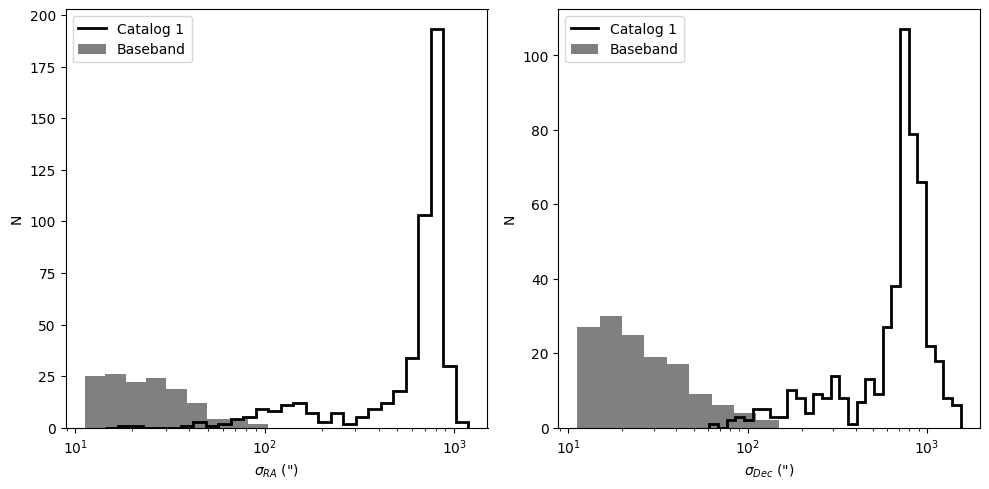}
\caption{\label{fig:loc}
  Distributions of 1-$\sigma$ uncertainties on RA (left) and Dec (right) reported in Catalog~1 and refined with baseband, respectively. 
  Multiple islands with similar probability are present in Catalog~1 localizations; only statistical uncertainties on the central island are reported in the plot.
  We note that full localization uncertainty contours are included in the Catalog~1 data release.
}
\end{figure*}

\subsubsection{Localization}
Localization regions obtained with baseband data are reported in Table~\ref{tab:results}.
They come in the form of ellipses on the sky, with axes nearly parallel to RA and Dec. 
Baseband localization regions are significantly smaller than those reported in Catalog~1, as shown in Fig.~\ref{fig:loc}.
Moreover, baseband data can, generally, be used to identify the correct beam's lobe pointing to the source position, removing the different `islands' of high probability visible in the localization contours of Catalog~1.
However, for a few events, the bandwidth and S/N were not large enough to robustly reject adjacent lobes.
For these FRBs, marked in Table~\ref{tab:results}, we report the most probable position but caution the reader that the localization is uncertain and their RA might be shifted up to $\pm3$\degr of arc.
We compared the baseband localization to the localization uncertainty contours calculated in Catalog~1 and found that $\sim 75\%$ of the positions are within the 1-$\sigma$ Catalog~1 contours and $\sim 95\%$ within the 2-$\sigma$ Catalog~1 contours.

\begin{figure}
\plotone{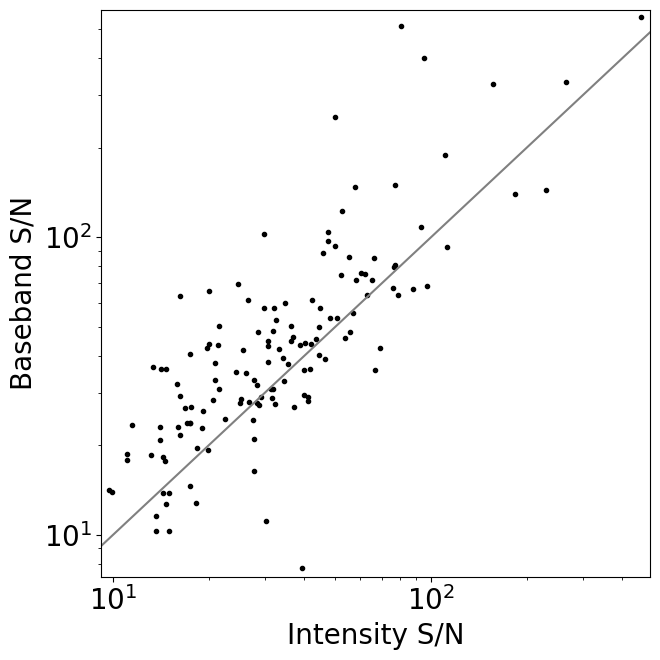}
\caption{\label{fig:snr}
  Comparison between S/N values obtained with baseband and intensity datasets. 
  Baseband values have been obtained as the total S/N of the main peak, while intensity values have been obtained with a least-squares fit.
  The gray line highlights where the two values are equal.
}
\end{figure}

\subsubsection{S/N}
The S/N values are compared with those reported in Catalog~1 in Fig.~\ref{fig:snr}.
Baseband S/N values are calculated for the region of the main peak in the pulse profile, while intensity values were obtained by using a least-squares fit implemented in \texttt{fitburst} \citep{chimefrbcatalog1,fpb+23}.
Therefore, the comparison is not completely accurate given the difference in S/N definitions; for example, baseband data would yield higher values if selecting the FWHM of the main peak instead of its full extent. 
S/N values calculated by applying \texttt{fitburst} to baseband data will be presented by \SandT.
In the meantime, the comparison is still interesting to check, for example, for potential biases in the analysis. 
The plot shows good general agreement between the two datasets.
Since, with baseband data, it is possible to beamform in the direction of the source, they will produce, on average, higher S/N values than intensity data, for which formed beams are static and the source could fall in a region of lower sensitivity. 
On the other hand, a fraction of frequency channels is often lost in a baseband dump due to a variety of system issues. 
The events where the baseband S/N is significantly lower than intensity S/N have been visually inspected, finding that indeed only a small fraction of the band had been stored in the baseband data.

\begin{figure}
\plotone{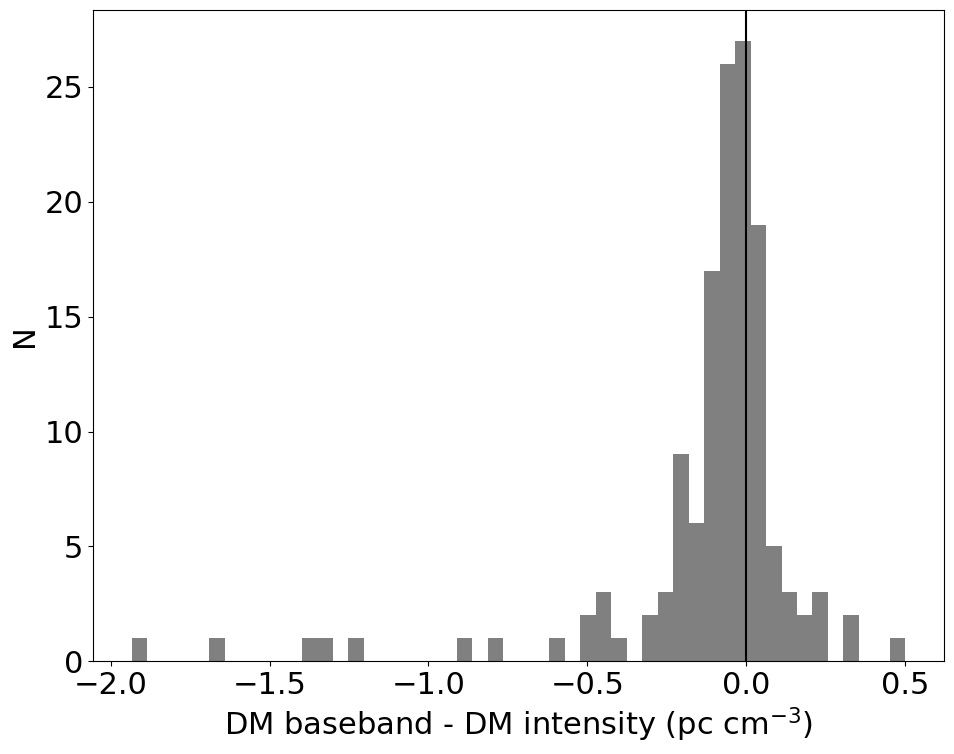}
\caption{\label{fig:dm_diff}
  Histogram of the difference between DM values measured with higher resolution baseband data and lower resolution intensity data. 
  The vertical line highlights identical values.
}
\end{figure}

\subsubsection{DM}
DM values can be measured more precisely with baseband data than with intensity data, as discussed in \textsection\ref{sec:intro}.
Also, the correlation between DM and exponential tails due to scattering can be attenuated with baseband data, because the differing frequency dependence becomes more obvious.
Finally, some FRBs show marching-down structures in their waterfall plots that can mimic the effect of dispersion \citep{hss+19}; this effect can also be reduced with the use of baseband data.
Since these last two effects are positively correlated with DM, DM values measured with a coarse time resolution tend to be, on average, larger than the actual values \citep{hss+19}.
This is noticeable in Fig.~\ref{fig:dm_diff}, where a comparison between the DM values obtained in Catalog~1 through the \texttt{fitburst} tool \citep{chimefrbcatalog1,fpb+23} and those obtained with baseband data through the \texttt{DM\_phase} algorithm \citep[][see \textsection\ref{sec:obs:dm}]{smp19} is reported.
Conversely, events with baseband DM larger than intensity DM, and inconsistent within uncertainties, have been visually inspected.
It is found that this is due to underestimated uncertainties on intensity values, sometimes related to the off-axis beam response that made part of the spectrum undetected.
A new version of \texttt{fitburst} is now available that more accurately calculates parameter uncertainties, which results in better statistical agreement between intensity and baseband DMs \citep{fpb+23}.

\begin{figure*}
\plotone{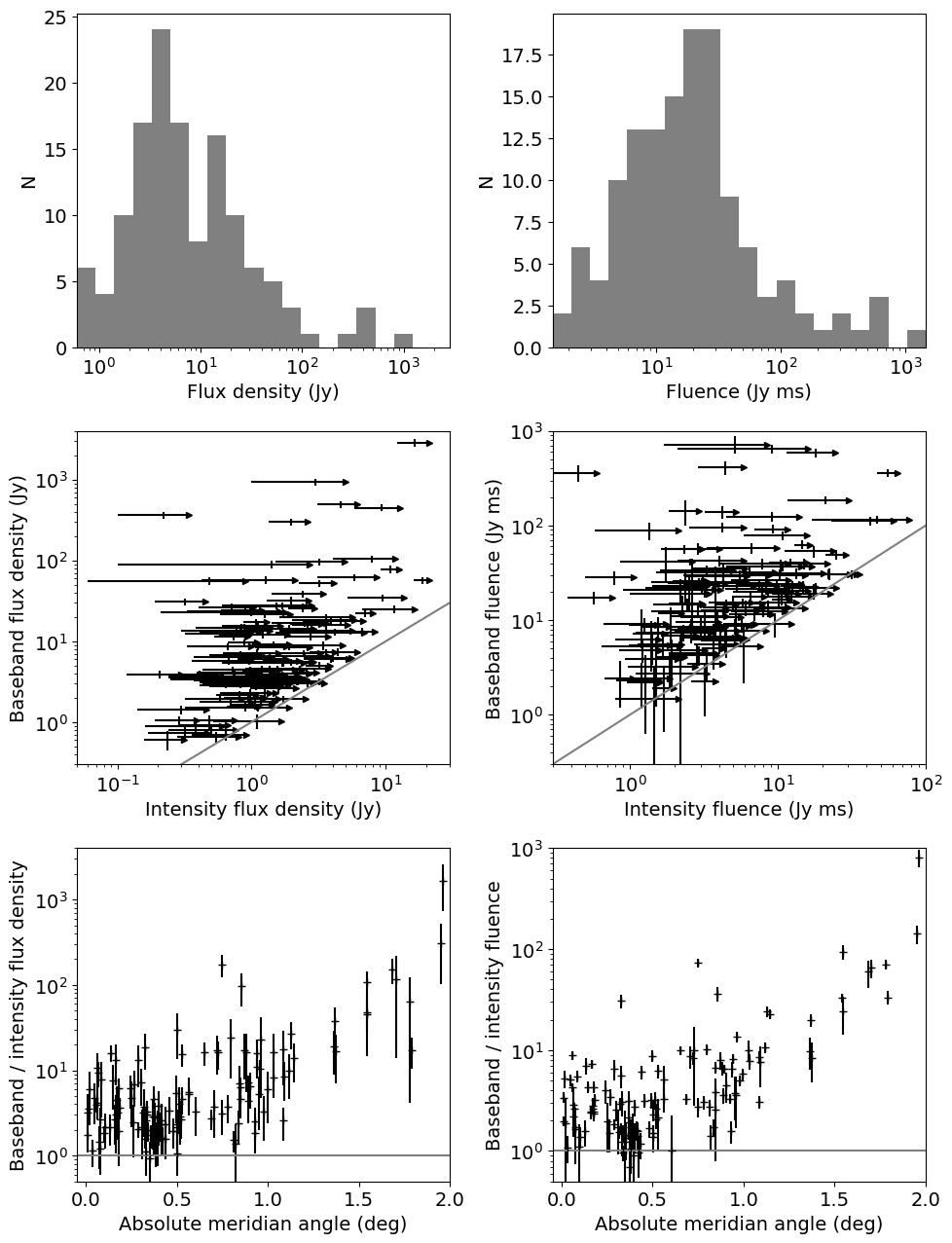}
\caption{\label{fig:flux}
  Measurements of flux densities (left) and fluences (right).
  Top: Histograms of values measured with baseband data.
  Middle: Comparison with Catalog~1. Values reported in Catalog~1 are lower limits, as highlighted by the arrows.
  Bottom: Ratio between baseband and intensity measurements as a function of the absolute meridian angle of the FRBs.
  Gray lines highlight identical values measured with the two datasets.
}
\end{figure*}

\subsubsection{Flux density and fluence}
The distribution of flux densities and fluences measured with baseband data is reported in Fig.~\ref{fig:flux} (top panel).
Fig.~\ref{fig:flux} also shows a comparison between intensity and baseband measurements of flux densities and fluences (middle panel).
As expected, values obtained with baseband data are systematically higher than those obtained with intensity data.
In the bottom panel, it is noticeable that the ratio between the two measurements tends to increase farther away from the telescope's meridian due to the effect of the primary beam.
The response of formed beams, on the other hand, causes the large scatter in the ratio since their sensitivity varies on much smaller scales than the primary beam, as discussed in \textsection\ref{sec:obs:flux} (see also \citealp{apb+23} for more details).

\subsection{Further studies of the population}
\label{sec:further_studies}
A series of papers is in preparation to analyze the properties of the 140 FRBs in detail.
These studies include a morphological analysis, which will be compared to a population of repeating FRBs; a study of the FRB scintillation, including using it to constrain the source distance; a cross-correlation of the FRB positions with large-scale structures in the Universe; and a comprehensive search for microlensing in the FRB signals.
Host galaxies of four FRBs in the sample have been presented by \citet{bmk+23}; analysis is in progress to extend the number of host associations in the local Universe.
Finally, the polarization properties of the FRB sample have been recently presented by \citet{ppm+24}, and they will be compared with the polarization properties of a large sample of repeating FRBs.

\section{Data release} \label{sec:data_release}
The results of our analysis for the sample of 140 FRBs are summarized in Table~\ref{tab:results}.
A machine-readable version of the same table can be downloaded from the online version of the manuscript.
Finally, a live online table will be updated as the new analyses described in \textsection\ref{sec:further_studies} are published.\footnote{\url{https://www.chime-frb.ca/baseband-catalog-1}}

In addition, we share the baseband data beamformed to the source position, as output from the single-beam stage of the baseband pipeline described in \textsection\ref{sec:obs:pipeline}.
All of the 140 beamformed files are released in the Hierarchical Data Format, version 5 (HDF5)\footnote{\url{www.hdfgroup.org}} and contain different arrays of data.
The main one consists of baseband data as a function of frequency, polarization, and time, stored at the instrumental resolution of 2.56 $\mu$s samples in 1024 frequency channels over 400 MHz.
Baseband data are calibrated and beamformed to the FRB position, but are not corrected for the source's dispersion.
To simplify data analysis, a second array is contained in the file with the power of the data obtained from the previous array, after dispersion smearing had been coherently removed.
Other arrays in the file map the central frequency of each channel, the UNIX time \citep{IEEE17} beginning of each channel, the sky position of the beam center, and the normalization factor to rescale the signal amplitude to units of Janskys.
Also, the HDF5 object and all the arrays contain metadata about stored data, such as the times of observation and processing, the pipeline version used, the time resolution of the data, the DM used for de-dispersion, etc.
The full content of the beamformed files is described in a README that is included in the data release.
Finally, a detailed example is provided in Python language showing how to use the arrays in the HDF5 files to produce a waterfall plot and calculate fluences.

The released data are assigned a Digital Object Identifier (DOI) and stored in the CANFAR Data Publication Service (DPS).\footnote{\url{https://doi.org/10.11570/23.0029}}
Finally, the CHIME/FRB team maintains a public page where the community can ask questions, provide feedback, and request additional information.\footnote{\url{https://github.com/chime-frb-open-data/community/discussions}}

\section{Conclusions}
We have processed the channelized raw voltages (baseband data) from each of the 1024 dual-polarization antennas of CHIME for 140 FRBs.
The properties of these sources have been previously reported in CHIME/FRB's Catalog 1 \citep{aab+21} by using downsampled total intensity data.
Here, we use baseband data to improve the measurements of the burst properties, including their position, DM, flux, fluence, and exposure.
A series of studies is currently in preparation to explore further properties of the sample enabled by the baseband catalog, including their polarization properties, scintillation, morphology, host identification, correlation with large-scale structures in the Universe, and microlensing in the FRB signals.
The baseband data for the sample of 140 FRBs, calibrated and beamformed to the FRB positions, is released to the community.

The CHIME/FRB experiment now routinely stores baseband data for every FRB detected with a $\text{S/N}>12$.
Efforts are ongoing to lower this threshold by quickly discarding data for false positives.
We are working to prepare a second release of baseband data for the hundreds of FRBs currently stored. 

\begin{acknowledgments}
We acknowledge that CHIME is located on the traditional, ancestral, and unceded territory of the Sylix (Okanagan) people.

We thank the Dominion Radio Astrophysical Observatory, operated by the National
Research Council Canada, for gracious hospitality and expertise. 
CHIME is funded by a grant from the Canada Foundation for Innovation (CFI) 2012 Leading Edge Fund (Project 31170) and by contributions from the provinces of British Columbia, Qu\'ebec and Ontario. 
The CHIME/FRB Project is funded by a grant from the CFI 2015 Innovation Fund (Project 33213) and by contributions from the provinces of British Columbia and Qu\'ebec, and by the Dunlap Institute for Astronomy and Astrophysics at the University of Toronto. Additional support was provided by the Canadian Institute for Advanced Research (CIFAR), McGill University and the McGill Space Institute via the Trottier Family Foundation, and the University of British Columbia. 
The Dunlap Institute is funded through an endowment established by the David Dunlap family and the University of Toronto. 
Research at Perimeter Institute is supported by the Government of Canada through Industry Canada and by the Province of Ontario through the Ministry of Research \& Innovation. 
The National Radio Astronomy Observatory is a facility of the National Science Foundation (NSF) operated under cooperative agreement by Associated Universities, Inc.  
FRB research at WVU is supported by an NSF grant (2006548, 2018490).
FRB research at UBC is supported by an NSERC Discovery Grant and by the Canadian Institute for Advanced Research.  
The baseband instrument for CHIME/FRB is funded in part by a CFI John R. Evans Leaders Fund grant to IHS.

This research used the Canadian Advanced Network For Astronomy Research (CANFAR) operated in partnership by the Canadian Astronomy Data Centre and The Digital Research Alliance of Canada with support from the National Research Council of Canada the Canadian Space Agency, CANARIE and the Canadian Foundation for Innovation.
\end{acknowledgments}

\begin{acknowledgments}
\allacks
\end{acknowledgments}


\clearpage

\appendix
\restartappendixnumbering

\section{Summary of results}
In Table~\ref{tab:results}, we report the results obtained with the methods described in \textsection\ref{sec:results}. 
An online version of this table is available with updated results.\footnote{\url{https://www.chime-frb.ca/baseband-catalog-1}}
Some exposures have zero uncertainties due to their definition and the precision of the source localization, as discussed in \textsection\ref{sec:obs:exposure}.

\startlongtable
\begin{deluxetable*}{lrlrllllll}
\tabletypesize{\footnotesize}
\tablecaption{\label{tab:results}
  Results of the analysis of 140 FRBs with baseband data ordered by time of detection. A machine-readable version of the table can be downloaded from the online version of the manuscript, while an online live table is maintained by the CHIME/FRB Collaboration (see \textsection\ref{sec:data_release}).
}
\tablehead{
  \colhead{TNS name} & 
  \colhead{RA\tablenotemark{a}} &
  \colhead{$\sigma_{\text{RA}}$\tablenotemark{b}} &
  \colhead{Dec\tablenotemark{a}} &
  \colhead{$\sigma_{\text{Dec}}$\tablenotemark{b}} &
  \colhead{DM\tablenotemark{c}} & 
  \colhead{Fluence} &
  \colhead{Flux Density} &
  \colhead{Up.\ exp.\tablenotemark{d}} &
  \colhead{Low.\ exp.\tablenotemark{d}} \\
   & 
  \colhead{\degr} &
  \colhead{\arcsec} &
  \colhead{\degr} &
  \colhead{\arcsec} &
  \colhead{pc cm$^{-3}$} & 
  \colhead{Jy ms} &
  \colhead{Jy} &
  \colhead{hr} &
  \colhead{hr}
}
\startdata
FRB 20181209A & 98.4137 & 24 & 68.5794 & 13 & 328.59(1) & 11(1) & 17(2) & 38(2) & -- \\ 
FRB 20181213A & 127.8826 & 26 & 73.9033 & 18 & 678.69(1) & 8.2(9) & 6.3(7) & 31(4) & 35.8(8) \\ 
FRB 20181214C & 175.8997 & 31 & 60.0573 & 25 & 632.832(3) & 15(2) & 4.5(5) & 26(2) & -- \\ 
FRB 20181215B & 254.6078 & 24 & 47.5892 & 12 & 494.044(6) & 13(1) & 14(1) & 25.9(2) & -- \\ 
FRB 20181219C & 17.8866 & 30 & 14.2603 & 30 & 647.68(4) & 17(2) & 3.9(4) & 11.2(9) & -- \\ 
FRB 20181220A\tablenotemark{e} & 348.7085 & 27 & 48.3457 & 18 & 209.525(8) & -- & -- & 21.0(7) & -- \\ 
FRB 20181221A & 230.4724 & 15 & 25.8542 & 16 & 316.25(5) & 15(2) & 3.3(3) & 20.04(9) & -- \\ 
FRB 20181221B & 315.0145 & 12 & 80.9405 & 13 & 1394.86(1) & 34(3) & 23(2) & 73(4) & 48(5) \\ 
FRB 20181222E & 50.3806 & 20 & 87.1169 & 27 & 327.989(4) & 30(3) & 8.7(9) & 329(9) & 180(10) \\ 
FRB 20181223C & 180.9319 & 23 & 27.5498 & 26 & 112.45(1) & 4.4(5) & 2.9(3) & 17.0(3) & -- \\ 
FRB 20181224E & 239.3002 & 15 & 7.2503 & 18 & 581.84(1) & 30(3) & 14(1) & 17.3(2) & -- \\ 
FRB 20181225A\tablenotemark{f} & 29.4629 & 24 & 65.7124 & 26 & 349.10(4) & 25(3) & 3.5(4) & 44.8(4) & -- \\ 
FRB 20181226A\tablenotemark{f} & 29.4894 & 19 & 65.7140 & 20 & 348.81(1) & 9(1) & 3.7(4) & 45.1(2) & -- \\ 
FRB 20181226D & 120.0764 & 13 & 22.1814 & 14 & 385.338(5) & 8.4(9) & 8.5(9) & 19(0) & -- \\ 
FRB 20181226E & 297.3697 & 19 & 73.7039 & 22 & 308.78(1) & 88(9) & 56(6) & 59(1) & 45.6(6) \\ 
FRB 20181228B\tablenotemark{e} & 254.6073 & 32 & 63.736 & 36 & 568.538(6) & -- & -- & 31(2) & -- \\ 
FRB 20181229A & 137.1914 & 26 & 41.9665 & 28 & 955.45(2) & 5.6(6) & 1.9(2) & 22.9(3) & -- \\ 
FRB 20181231A & 29.672 & 72 & 21.010 & 100 & 1376.9(3) & 2.7(4) & 1.0(2) & 16(2) & -- \\ 
FRB 20181231B & 130.8028 & 24 & 56.0239 & 13 & 197.366(9) & 56(6) & 24(2) & 31.4(2) & -- \\ 
FRB 20181231C\tablenotemark{g} & 202.332 & 49 & 69.214 & 42 & 556.03(2) & 13(1) & 7.2(9) & 47(1) & -- \\ 
FRB 20190102A & 7.5613 & 18 & 26.7447 & 22 & 699.1(4) & 10(1)0 & 15(2) & 15.8(5) & -- \\ 
FRB 20190102B & 21.7419 & 20 & 21.5075 & 23 & 367.07(4) & 11(1) & 5.4(6) & 14.5(7) & -- \\ 
FRB 20190103C & 104.0400 & 25 & 11.0347 & 34 & 1349.3(1) & 31(3) & 4.5(5) & 7(2) & -- \\ 
FRB 20190106B & 338.0177 & 24 & 46.1942 & 13 & 316.536(2) & 25(3) & 27(3) & 21.9(6) & -- \\ 
FRB 20190110A & 67.4579 & 13 & 47.4804 & 14 & 472.788(3) & 32(3) & 27(3) & 12(3) & -- \\ 
FRB 20190110C & 249.3278 & 25 & 41.4434 & 26 & 222.01(1) & 5.3(6) & 3.4(4) & 14(2) & -- \\ 
FRB 20190111B & 259.9929 & 14 & 13.5434 & 17 & 1336.87(1) & 28(3) & 31(3) & 19(0) & -- \\ 
FRB 20190115B & 77.5418 & 25 & 82.0083 & 29 & 748.18(3) & 7.8(9) & 3.5(4) & 128(2) & 127.9(9) \\ 
FRB 20190116A\tablenotemark{h} & 192.270 & 66 & 27.164 & 88 & 445.0(5) & 6.5(8) & 0.7(1) & 19.9(3) & -- \\ 
FRB 20190117A\tablenotemark{i} & 331.6579 & 15 & 17.3688 & 16 & 393.13(5) & 24(2) & 6.6(7) & 18.7(1) & -- \\ 
FRB 20190118A & 255.1068 & 11 & 11.4984 & 12 & 225.108(5) & 590(60) & 450(40) & 12.1(5) & -- \\ 
FRB 20190121A & 354.5985 & 14 & 78.5798 & 15 & 425.28(3) & 37(4) & 5.4(6) & 69(2) & 4(2) \\ 
FRB 20190122C & 200.4987 & 12 & 17.5914 & 13 & 690.032(8) & 120(10) & 13(1) & 19(0) & -- \\ 
FRB 20190124B & 214.7106 & 30 & 28.774 & 36 & 441.6(2) & 20(2) & 2.3(3) & 17.1(2) & -- \\ 
FRB 20190124F & 339.0160 & 14 & 5.3271 & 17 & 254.799(4) & 32(3) & 18(2) & 16(0) & -- \\ 
FRB 20190130B & 173.7121 & 14 & 16.0527 & 15 & 988.75(1) & 24(3) & 13(1) & 12.8(8) & -- \\ 
FRB 20190131E & 206.7288 & 14 & 80.7734 & 18 & 279.798(6) & 720(70) & 930(90) & 109.7(9) & 92(2) \\ 
FRB 20190201B & 118.3688 & 21 & 55.4638 & 22 & 749.07(2) & 7.4(8) & 3.3(4) & 22(2) & -- \\ 
FRB 20190202B\tablenotemark{e,j} & 105.5490 & 24 & 31.9041 & 14 & 464.839(4) & -- & -- & 12.0(9) & -- \\ 
FRB 20190203A & 128.9984 & 15 & 70.8421 & 16 & 420.586(6) & 42(4) & 12(1) & 53.7(3) & 38.8(2) \\ 
FRB 20190204B & 255.7251 & 35 & 75.9264 & 35 & 1464.842(6) & 6.2(7) & 3.1(3) & 70(2) & 44(2) \\ 
FRB 20190206A & 246.2702 & 17 & 9.3321 & 21 & 188.353(3) & 640(60) & 89(9) & 18.3(1) & -- \\ 
FRB 20190208C & 126.0334 & 17 & 83.4001 & 20 & 238.323(5) & 42(5) & 58(6) & 118(4) & 75(8) \\ 
FRB 20190210B & 104.0372 & 13 & 23.7992 & 15 & 624.24(1) & 23(2) & 16(2) & 16.1(5) & -- \\ 
FRB 20190212B & 139.7677 & 18 & 52.2019 & 19 & 600.185(3) & 7.7(8) & 4.4(5) & 24(1) & -- \\ 
FRB 20190212C & 172.4552 & 32 & 28.146 & 38 & 1015.6(7) & 40(4) & 1.8(2) & 7(3) & -- \\ 
FRB 20190213D & 338.498 & 45 & 52.748 & 76 & 1346.7(4) & 19(2) & 3.7(5) & 28.1(8) & -- \\ 
FRB 20190214C & 218.8294 & 20 & 19.2882 & 26 & 532.95(1) & 13(1) & 3.2(3) & 17.9(3) & -- \\ 
FRB 20190217A & 94.8603 & 33 & 43.246 & 43 & 798.14(4) & 6.2(7) & 1.0(1) & 12(4) & -- \\ 
FRB 20190224C & 125.1238 & 19 & 19.8146 & 22 & 497.12(2) & 39(4) & 4.5(5) & 19.2(2) & -- \\ 
FRB 20190224D & 338.0643 & 16 & 89.0782 & 19 & 752.892(6) & 13(1) & 11(1) & 1280(20) & 1100(20) \\ 
FRB 20190226A & 58.8369 & 18 & 31.9120 & 20 & 601.546(7) & 5.4(6) & 9.1(9) & 17.2(2) & -- \\ 
FRB 20190227A & 108.1239 & 14 & 56.2749 & 14 & 394.031(8) & 23(2) & 7.3(7) & 29.8(6) & -- \\ 
FRB 20190301A\tablenotemark{k} & 313.0391 & 19 & 69.7435 & 21 & 459.44(2) & 14(2) & 3.3(4) & 41(1) & -- \\ 
FRB 20190303B & 128.6877 & 11 & 66.0059 & 11 & 193.429(5) & 110(10) & 34(3) & 32(2) & -- \\ 
FRB 20190304A & 118.3675 & 26 & 74.6021 & 18 & 483.521(8) & 57(6) & 27(3) & 68.2(2) & 36(2) \\ 
FRB 20190304B & 204.9119 & 35 & 24.0913 & 34 & 469.90(2) & 6.9(8) & 3.1(4) & 16.8(5) & -- \\ 
FRB 20190320A & 61.3282 & 34 & 63.361 & 41 & 614.2(1) & 6.1(7) & 2.0(2) & 39.9(2) & -- \\ 
FRB 20190320B & 250.5549 & 15 & 39.7956 & 16 & 489.501(8) & 12(1) & 11(1) & 23(0) & -- \\ 
FRB 20190320E & 67.7107 & 19 & 89.1340 & 22 & 299.09(2) & 18(2) & 7.3(8) & 1280(20) & 1120(20) \\ 
FRB 20190322A & 110.564 & 93 & 51.176 & 78 & 1059.68(8) & 15(2) & 0.69(9) & 19(4) & -- \\ 
FRB 20190323B & 192.8926 & 12 & 77.1408 & 12 & 789.527(7) & 16(2) & 22(2) & 65(2) & 74.5(1) \\ 
FRB 20190327A & 281.3318 & 14 & 34.2765 & 14 & 346.579(7) & 8.9(9) & 7.0(7) & 15.3(8) & -- \\ 
FRB 20190405B & 306.6506 & 33 & 88.6207 & 28 & 1113.72(7) & 53(6) & 9(1) & 682(8) & 685(3) \\ 
FRB 20190410B & 267.6038 & 14 & 15.1121 & 17 & 642.152(8) & 360(40) & 360(40) & 18.3(2) & -- \\ 
FRB 20190411B & 154.102 & 48 & 29.317 & 119 & 1229.417(7) & 5.0(6) & 1.5(2) & 15(1) & -- \\ 
FRB 20190411C & 10.6856 & 11 & 20.4790 & 11 & 233.714(8) & 92(9) & 52(5) & 19.20(9) & -- \\ 
FRB 20190412A & 243.2154 & 18 & 61.8405 & 19 & 364.55(1) & 12(1) & 3.5(4) & 37.9(2) & -- \\ 
FRB 20190417C & 45.6485 & 11 & 71.2554 & 11 & 320.266(4) & 78(8) & 100(10) & 10(9) & 39.7(5) \\ 
FRB 20190418A & 65.8093 & 26 & 16.0791 & 33 & 184.473(3) & 4.8(6) & 2.9(3) & 18.4(3) & -- \\ 
FRB 20190419B & 256.4518 & 15 & 86.7633 & 19 & 165.13(2) & 21(2) & 18(2) & 303(3) & 40(3) \\ 
FRB 20190420B & 94.443 & 38 & 70.074 & 39 & 846.646(9) & 40(4) & 3.2(4) & 21(9) & 41(0) \\ 
FRB 20190423A & 179.4682 & 11 & 55.3180 & 11 & 242.600(8) & 360(40) & 78(8) & 28.2(6) & -- \\ 
FRB 20190423D & 30.0391 & 23 & 84.6806 & 30 & 496(1) & 22(2) & 1.9(2) & 175(5) & 190(2) \\ 
FRB 20190425A & 255.6698 & 11 & 21.5779 & 11 & 128.14(1) & 30(3) & 57(6) & 18.4(2) & -- \\ 
FRB 20190425B & 153.4102 & 15 & 88.3676 & 17 & 1031.63(1) & 25(3) & 22(2) & 541(5) & 270(30) \\ 
FRB 20190427A & 78.9486 & 22 & 7.8184 & 29 & 455.78(1) & 9(1) & 6.1(7) & 17.8(2) & -- \\ 
FRB 20190430C & 277.2066 & 15 & 24.7650 & 16 & 400.3(3) & 8.6(9) & 4.0(4) & 8(2) & -- \\ 
FRB 20190501B & 261.1994 & 27 & 54.2544 & 19 & 783.967(4) & 21(2) & 10(1) & 20(2) & -- \\ 
FRB 20190502A & 165.0863 & 15 & 59.9156 & 16 & 625.74(1) & 19(2) & 5.0(5) & 35.4(2) & -- \\ 
FRB 20190502B & 212.0159 & 33 & 64.459 & 38 & 918.6(2) & 20(2) & 4.1(5) & 11(7) & -- \\ 
FRB 20190502C & 155.1462 & 17 & 82.9981 & 25 & 396.878(9) & 26(3) & 20(2) & 124(4) & 18(3) \\ 
FRB 20190517C\tablenotemark{e} & 89.4981 & 13 & 26.5541 & 14 & 335.54(6) & -- & -- & 18.3(3) & -- \\ 
FRB 20190518C & 241.9646 & 16 & 4.6332 & 21 & 443.964(6) & 22(2) & 13(1) & 17.3(2) & -- \\ 
FRB 20190519E & 169.627 & 60 & 41.709 & 79 & 693.622(7) & 1.5(2) & 3.3(4) & 21(1) & -- \\ 
FRB 20190519G & 306.6143 & 27 & 72.4144 & 30 & 429.5(5) & 31(3) & 1.5(2) & 56.0(9) & 17(3) \\ 
FRB 20190519H & 339.8638 & 12 & 87.3777 & 12 & 1170.878(6) & 58(6) & 10(1)0 & 180(10) & 381(3) \\ 
FRB 20190604G & 120.8081 & 15 & 59.5008 & 16 & 232.998(7) & 15(1) & 4.2(4) & 36.1(3) & -- \\ 
FRB 20190606A\tablenotemark{f} & 29.5339 & 17 & 65.7161 & 20 & 349.8(5) & 24(2) & 3.8(4) & 45.1(2) & -- \\ 
FRB 20190606B\tablenotemark{f} & 29.4852 & 16 & 65.7143 & 17 & 348.96(3) & 49(5) & 3.2(3) & 45.1(2) & -- \\ 
FRB 20190605C & 170.1120 & 12 & -5.1578 & 15 & 187.713(5) & 410(40) & 490(50) & 17(0) & -- \\ 
FRB 20190606A\tablenotemark{l} & 218.714 & 62 & 53.313 & 58 & 552.552(8) & 3.2(4) & 1.0(2) & 29.4(5) & -- \\ 
FRB 20190606B & 108.8065 & 25 & 86.8053 & 31 & 277.67(3) & 19(2) & 3.0(3) & 320(3) & 55(3) \\ 
FRB 20190607B & 41.936 & 37 & 49.624 & 40 & 289.331(2) & 2.2(3) & 1.6(2) & 28.2(4) & -- \\ 
FRB 20190608A & 359.226 & 43 & 19.175 & 49 & 722.14(1) & 3.9(5) & 6.0(7) & 18.8(1) & -- \\ 
FRB 20190609A & 21.8256 & 26 & 87.7004 & 23 & 316.684(3) & 37(4) & 16(2) & 446(5) & 98(4) \\ 
FRB 20190609B & 208.3177 & 24 & 88.3412 & 15 & 292.174(7) & 31(3) & 25(3) & 552(0) & 348(0) \\ 
FRB 20190609C & 73.3278 & 35 & 24.098 & 37 & 479.852(5) & 4.1(5) & 3.0(4) & 19.5(3) & -- \\ 
FRB 20190609D & 118.513 & 39 & 51.742 & 40 & 511.56(2) & 23(3) & 12(1) & 27.8(7) & -- \\ 
FRB 20190611A\tablenotemark{m} & 65.6706 & 23 & 73.6545 & 27 & 191(2) & 23(2) & 1.8(2) & 64.4(3) & 47.5(3) \\ 
FRB 20190612A\tablenotemark{n} & 148.0969 & 31 & 70.520 & 36 & 433.14 & -- & -- & 42(3) & 26(2) \\ 
FRB 20190612B & 222.1114 & 16 & 4.3905 & 23 & 187.524(7) & 14(1) & 25(3) & 16(0) & -- \\ 
FRB 20190613A & 257.4063 & 29 & 18.9287 & 36 & 714.71(3) & 13(1) & 2.8(3) & 17.2(6) & -- \\ 
FRB 20190613B & 65.7428 & 13 & 42.6817 & 14 & 285.088(5) & 8.8(9) & 17(2) & 24.2(2) & -- \\ 
FRB 20190614A & 230.7554 & 22 & 88.0029 & 24 & 1063.917(6) & 22(2) & 13(1) & 311(5) & 410(10) \\ 
FRB 20190614C & 356.484 & 53 & 35.992 & 73 & 589.1(1) & 4.3(5) & 0.8(1) & 19(1) & -- \\ 
FRB 20190616A & 234.2437 & 20 & 34.4844 & 22 & 212.511(5) & 5.5(6) & 4.1(4) & 21.7(3) & -- \\ 
FRB 20190617A & 177.7378 & 11 & 83.8060 & 11 & 195.749(6) & 190(20) & 62(6) & 150(0) & 161.1(2) \\ 
FRB 20190617B & 58.0429 & 17 & 1.4016 & 29 & 272.73(7) & 120(10) & 23(2) & 17(0) & -- \\ 
FRB 20190617C & 134.289 & 46 & 35.742 & 50 & 638.90(2) & 6.7(7) & 0.6(1) & 20.9(6) & -- \\ 
FRB 20190618A & 323.1785 & 12 & 25.4747 & 12 & 228.920(6) & 35(4) & 38(4) & 17.0(3) & -- \\ 
FRB 20190619A & 165.1518 & 22 & 68.3139 & 25 & 899.82(1) & 3.5(4) & 2.6(3) & 46.1(6) & -- \\ 
FRB 20190619B & 231.5826 & 24 & 82.0046 & 28 & 270.549(3) & 6.4(7) & 5.5(6) & 86(7) & 127(1) \\ 
FRB 20190619C & 39.6424 & 17 & 36.2258 & 18 & 488.072(3) & 4.8(5) & 4.4(5) & 17(1) & -- \\ 
FRB 20190619D & 114.6547 & 28 & 41.7227 & 28 & 378.8(2) & 18(2) & 0.9(1) & 12(3) & -- \\ 
FRB 20190621A\tablenotemark{g,o} & 188.090 & 48 & 74.135 & 71 & 195.94(2) & 3.5(4) & 1.0(2) & 30(10) & 37(2) \\ 
FRB 20190621B & 192.838 & 68 & 55.621 & 71 & 1061.14(2) & 2.3(3) & 1.4(2) & 31.2(3) & -- \\ 
FRB 20190621C & 204.9242 & 17 & 5.2158 & 23 & 570.342(7) & 140(10) & 300(30) & 17.9(3) & -- \\ 
FRB 20190621D & 278.0152 & 21 & 78.8804 & 26 & 647.32(4) & 29(3) & 6.3(7) & 94.4(4) & 85.0(9) \\ 
FRB 20190622A & 298.467 & 57 & 85.781 & 62 & 1122.807(9) & 2.4(3) & 3.6(4) & 245.1(9) & 220(8) \\ 
FRB 20190623A & 270.5694 & 31 & 24.495 & 36 & 1082.16(1) & 2.4(3) & 3.9(4) & 20.0(2) & -- \\ 
FRB 20190623C & 193.502 & 47 & 86.045 & 58 & 1049.94(1) & 5.3(7) & 3.0(4) & 200(10) & 243(1) \\ 
FRB 20190624A & 168.0632 & 34 & 69.7932 & 30 & 973.9(1) & 10(1) & 2.0(2) & 43(0) & -- \\ 
FRB 20190624B & 308.6502 & 23 & 73.5968 & 11 & 213.947(8) & 1500(100) & 2900(300) & 63(0) & 49(0) \\ 
FRB 20190625E\tablenotemark{m} & 65.685 & 46 & 73.670 & 48 & 188.61(3) & 140(10) & 14(2) & 63.8(7) & 47.0(6) \\ 
FRB 20190626A\tablenotemark{e,m} & 65.701 & 37 & 73.669 & 37 & 191.2(5) & -- & -- & 64.0(4) & 47.1(3) \\ 
FRB 20190627A & 196.028 & 67 & 0.980 & 149 & 404.3(1) & 26(3) & 32(4) & 16(0) & -- \\ 
FRB 20190627C & 267.7766 & 25 & 71.6293 & 16 & 968.50(1) & 13(1) & 6.3(6) & 55.2(5) & 35.3(6) \\ 
FRB 20190627D & 295.431 & 104 & 43.856 & 91 & 2000.31(3) & 9(1) & 0.6(2) & 24.4(5) & -- \\ 
FRB 20190628A & 199.164 & 38 & 51.696 & 41 & 745.790(8) & 2.2(3) & 2.2(2) & 27.3(6) & -- \\ 
FRB 20190628B & 249.194 & 41 & 80.093 & 48 & 407.99(2) & 7.2(9) & 4.0(5) & 8(6) & 80(3) \\ 
FRB 20190628C\tablenotemark{g} & 11.453 & 101 & 48.591 & 99 & 1746.8(3) & -- & -- & 8(6) & -- \\ 
FRB 20190629A & 4.877 & 45 & 12.501 & 52 & 503.54(3) & 24(3) & 6.8(8) & 9(2) & -- \\ 
FRB 20190630B & 328.3149 & 13 & 42.9525 & 13 & 651.7(3) & 63(6) & 3.6(4) & 22.3(4) & -- \\ 
FRB 20190630C & 67.3687 & 22 & 80.9195 & 29 & 1660.21(1) & 15(2) & 8.6(9) & 10(4) & 81(3) \\ 
FRB 20190630D & 143.4744 & 19 & 8.8140 & 25 & 323.540(3) & 8.0(9) & 5.7(6) & 16.4(3) & -- \\ 
FRB 20190701A & 277.242 & 43 & 59.029 & 56 & 637.091(9) & 1.9(2) & 2.3(3) & 35.4(3) & -- \\ 
FRB 20190701B & 302.1952 & 20 & 80.0913 & 24 & 749.093(8) & 9(1) & 7.2(8) & 87(4) & 97(1) \\ 
FRB 20190701C & 83.7980 & 34 & 81.594 & 42 & 973.79(1) & 21(3) & 15(2) & 124.8(7) & 60(10) \\ 
FRB 20190701D & 111.7658 & 16 & 66.7667 & 17 & 933.32(3) & 20(2) & 3.1(3) & 39(1) & -- \\
\enddata
\tablenotetext{a}{J2000}
\tablenotetext{b}{Seconds of arc as distances on the sky}
\tablenotetext{c}{DM calculated with \texttt{DM\_phase} to maximize the burst structure}
\tablenotetext{d}{Exposures for upper and lower (when applicable) transits of the source}
\tablenotetext{e}{The absolute meridian angle is $>2$ deg, where the primary beam is unmodeled; flux and fluence are not calculated}
\tablenotetext{f}{Event of the repeating source FRB 20180916B \citep{abb+19c}}
\tablenotetext{g}{The correct lobe of the formed beam is uncertain, RA could be shifted by up to $\pm\frac{3}{\cos(\text{Dec})}$ deg of arc}
\tablenotetext{h}{Repeating source \citep{abb+19c}}
\tablenotetext{i}{Repeating source \citep{fab+20}}
\tablenotetext{j}{FRB detected in a sidelobe of the primary beam, where systematics are not constrained; its position might be affected by unaccounted systematic errors}
\tablenotetext{k}{Event of the repeating source FRB 20190222A \citep{abb+19c}}
\tablenotetext{l}{Event of the repeating source FRB 20190604A \citep{fab+20}}
\tablenotetext{m}{Event of the repeating source FRB 20180814A \citep{abb+19b}}
\tablenotetext{n}{A significant fraction of the signal is outside the time range of the dump; the DM was calculated by maximizing the S/N}
\tablenotetext{o}{Event of the repeating source FRB 20180908B \citep{fab+20}}
\end{deluxetable*}

\clearpage

\section{Waterfall plots}
Waterfall plots of the signal intensity as a function of frequency and time are presented in Fig.~\ref{fig:wfalls}.
Note that two `wings' are visible in the waterfall plot of very bright sources (e.g., FRB 20190417C).
These features are unphysical and arise from the channelization of the data.
The FRB spectra have not been corrected for the effect of the primary beam response of the telescope in the plots, while this effect has been removed in the flux and fluence calculations. This means that the telescope's sensitivity is larger at lower frequencies for bursts detected away from the meridian. Therefore, bursts may appear brighter at lower frequencies as a result of this effect. FRB spectra corrected for the primary beam response will be presented in a future work.

\begin{figure*}[b]
\epsscale{1.0}
\plotone{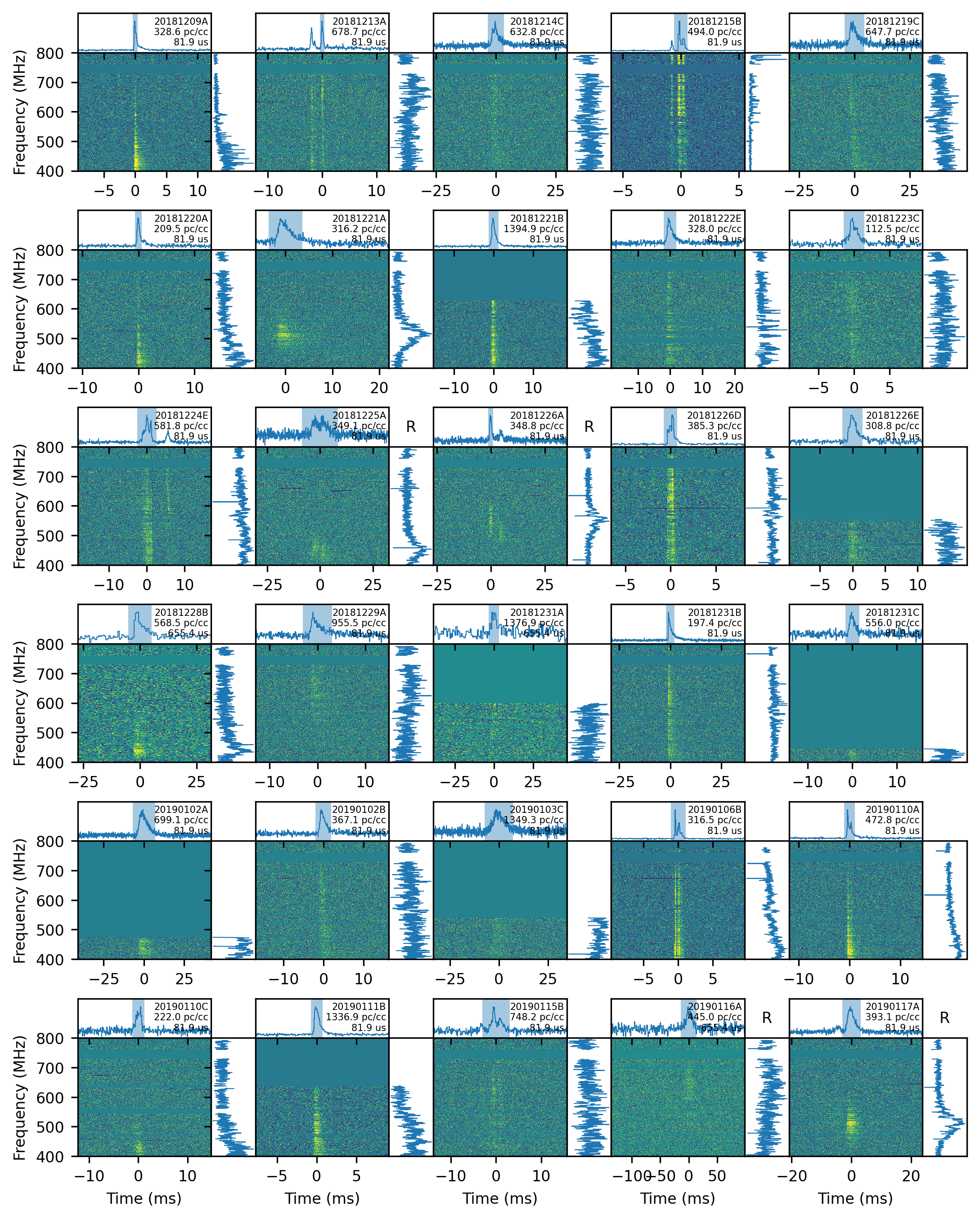}
\caption{\label{fig:wfalls}
  Waterfall plots of FRB signal intensities as a function of frequency and time. 
  Missing channels have been replaced with the median value of the waterfall.
  Pulse profiles are reported on top of each waterfall.
  Blue regions highlight the time ranges used to calculate the spectra plotted on the right of each waterfall.
  The TNS name, DM, and time resolution are reported in each subplot, while an `R' indicates bursts from a repeating source.
  The frequency resolution is $391$ kHz.
  FRB spectra have not been corrected for the response of the primary beam  of the telescope.
}
\end{figure*}

\setcounter{figure}{0}
\begin{figure*}
\epsscale{1.1}
\plotone{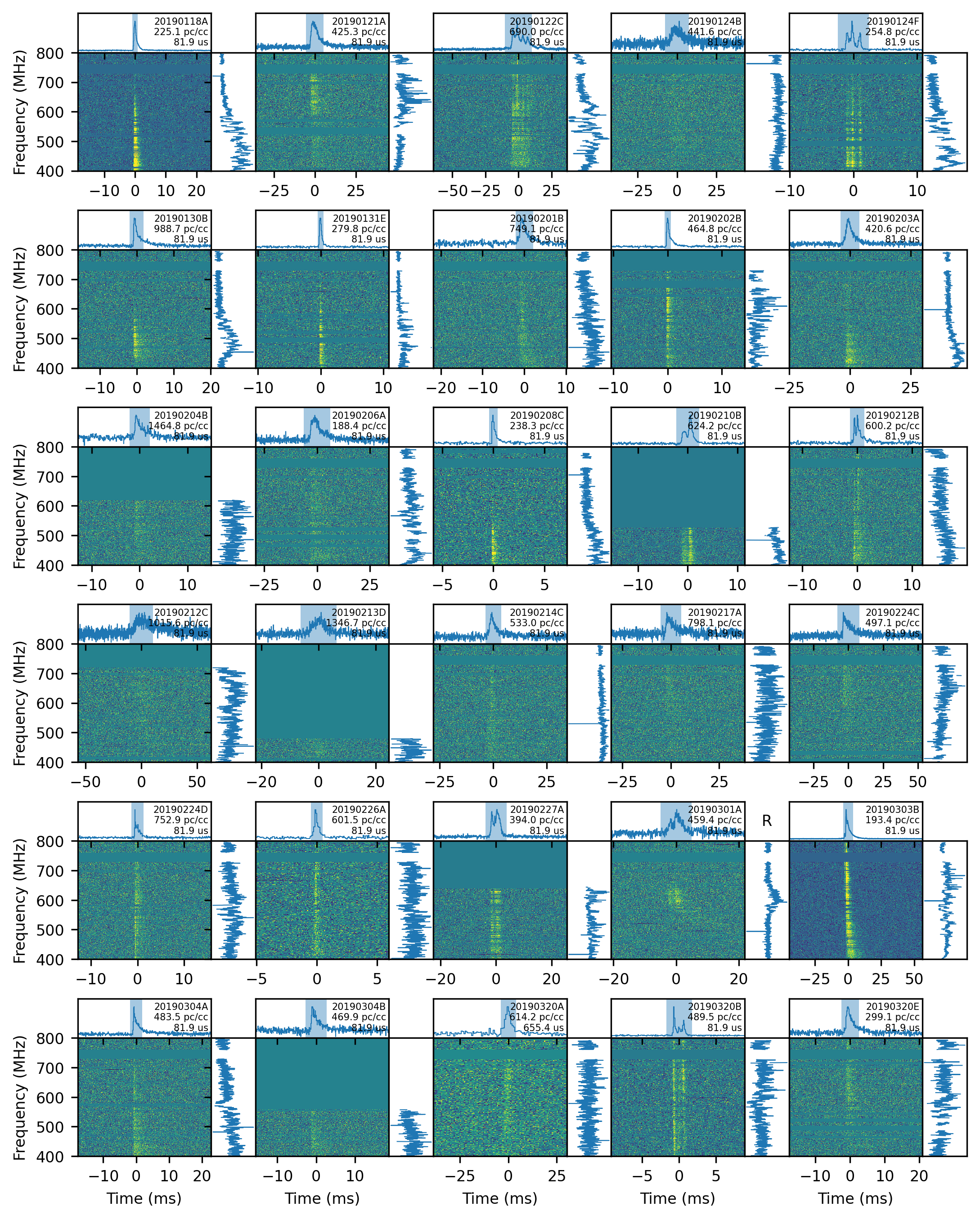}
\caption{Continued.}
\end{figure*}

\setcounter{figure}{0}
\begin{figure*}
\epsscale{1.1}
\plotone{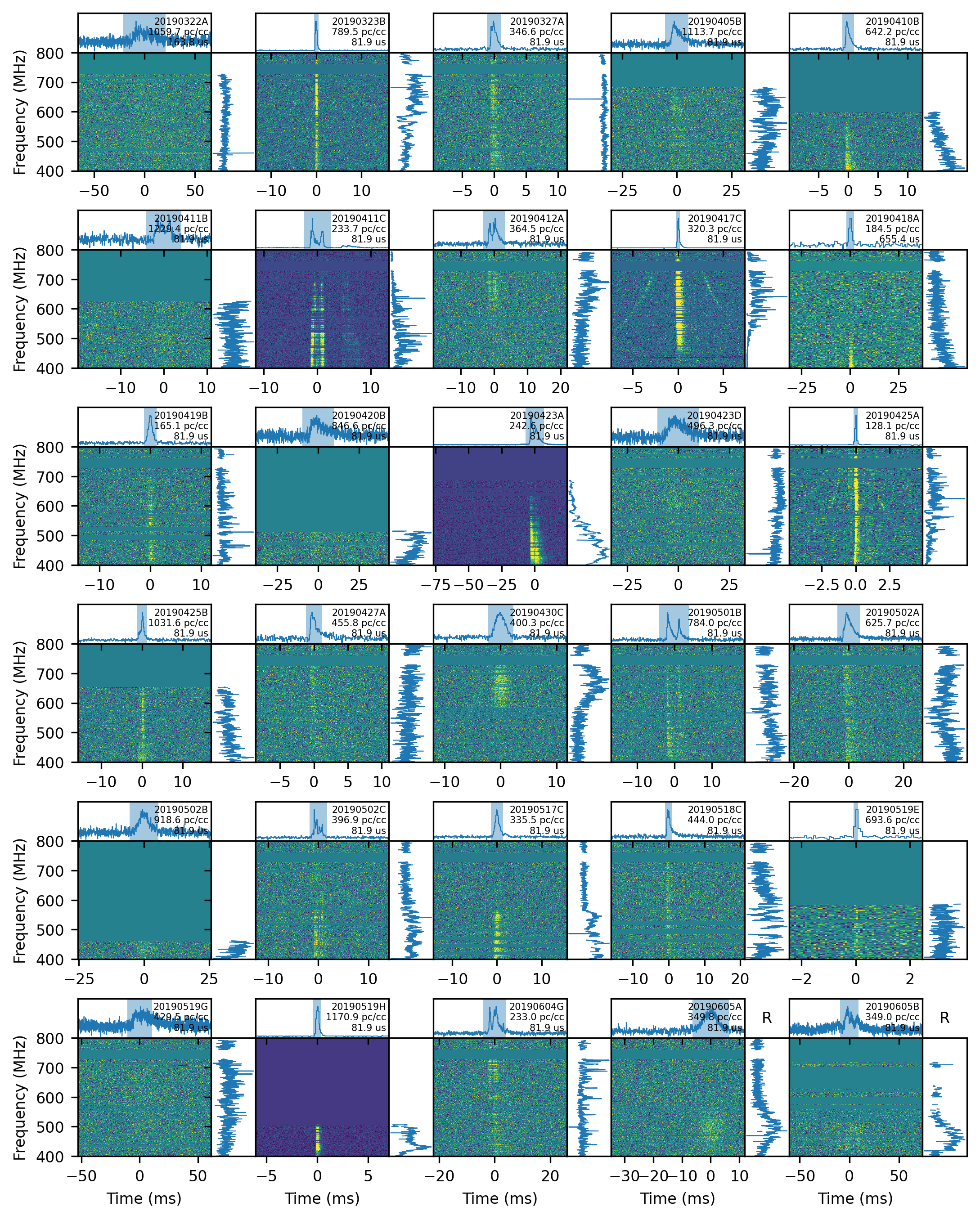}
\caption{Continued.}
\end{figure*}

\setcounter{figure}{0}
\begin{figure*}
\epsscale{1.1}
\plotone{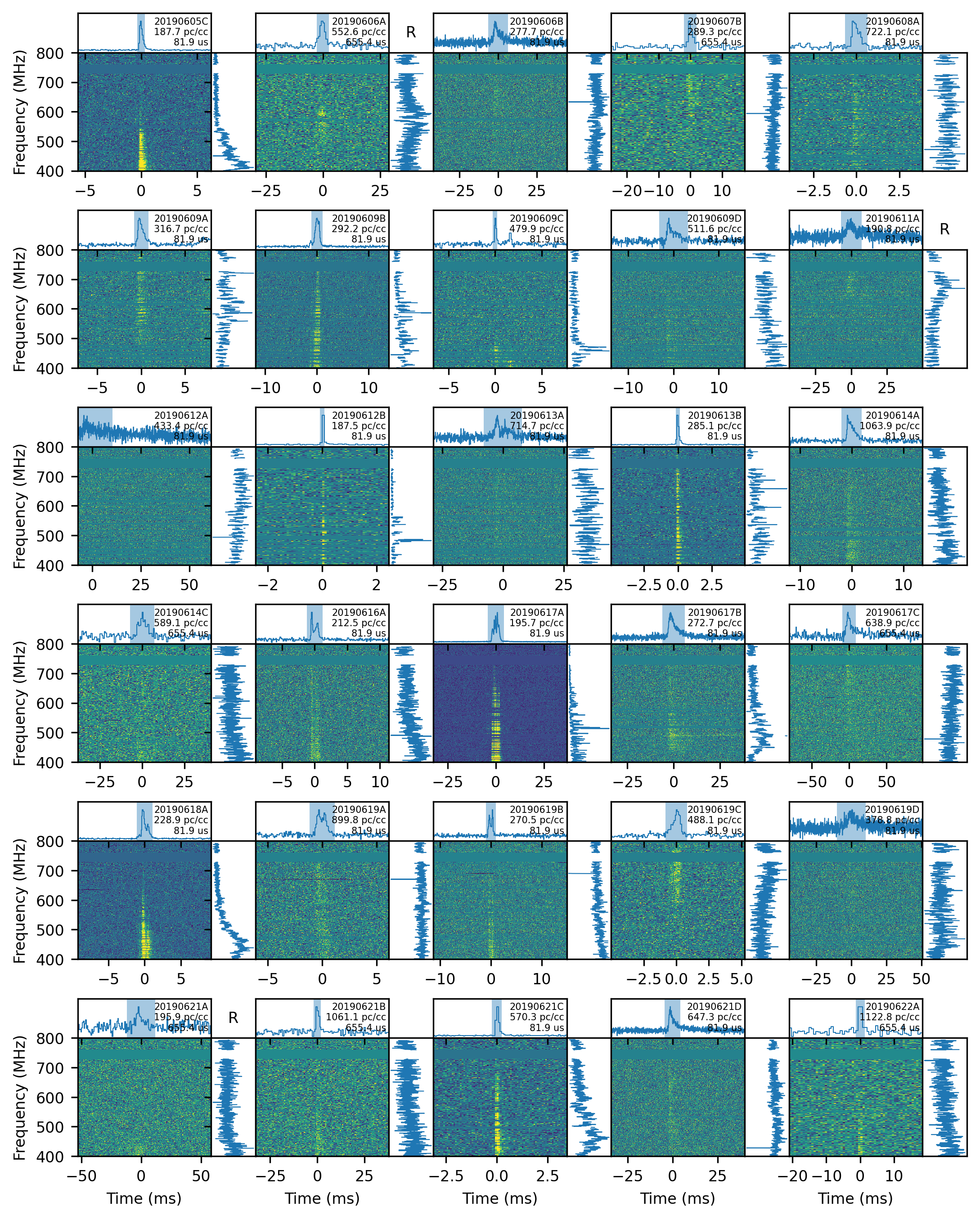}
\caption{Continued.}
\end{figure*}

\setcounter{figure}{0}
\begin{figure*}
\epsscale{1.1}
\plotone{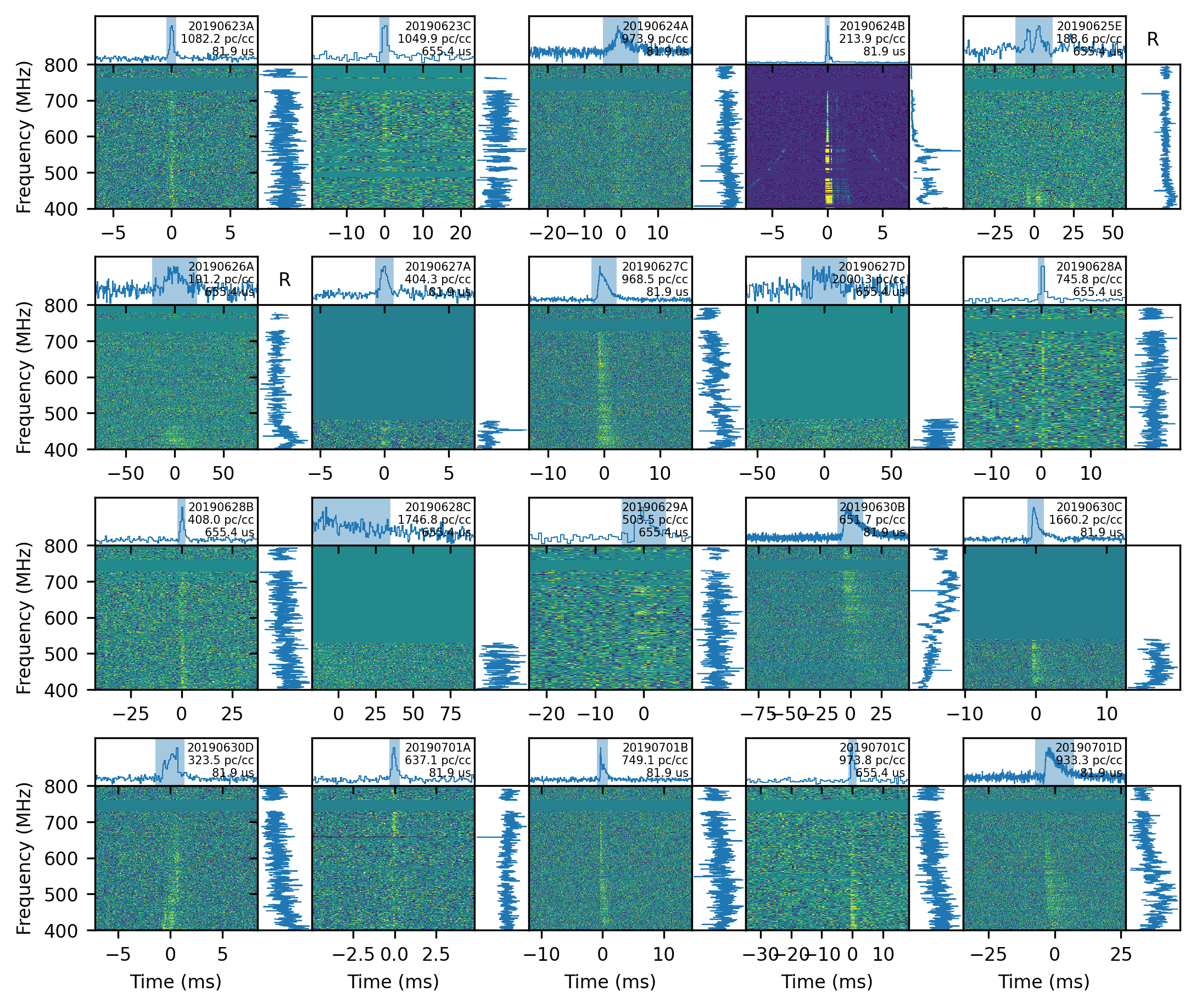}
\caption{Continued.}
\end{figure*}

\clearpage

\bibliography{frbrefs}{}
\bibliographystyle{aasjournal}

\end{document}

%% file: auth.tex
\author[0000-0001-6523-9029]{Mandana Amiri}
  \affiliation{Department of Physics and Astronomy, University of British Columbia, 6224 Agricultural Road, Vancouver, BC V6T 1Z1 Canada}
\author[0000-0001-5908-3152]{Bridget C.~Andersen}
  \affiliation{Department of Physics, McGill University, 3600 rue University, Montr\'eal, QC H3A 2T8, Canada}
  \affiliation{Trottier Space Institute, McGill University, 3550 rue University, Montr\'eal, QC H3A 2A7, Canada}
\author[0000-0002-3980-815X]{Shion Andrew}
  \affiliation{MIT Kavli Institute for Astrophysics and Space Research, Massachusetts Institute of Technology, 77 Massachusetts Ave, Cambridge, MA 02139, USA}
  \affiliation{Department of Physics, Massachusetts Institute of Technology, 77 Massachusetts Ave, Cambridge, MA 02139, USA}
\author[0000-0003-3772-2798]{Kevin Bandura}
  \affiliation{Lane Department of Computer Science and Electrical Engineering, 1220 Evansdale Drive, PO Box 6109 Morgantown, WV 26506, USA}
  \affiliation{Center for Gravitational Waves and Cosmology, West Virginia University, Chestnut Ridge Research Building, Morgantown, WV 26505, USA}
\author[0000-0002-3615-3514]{Mohit Bhardwaj}
  \affiliation{Department of Physics, Carnegie Mellon University, 5000 Forbes Avenue, Pittsburgh, 15213, PA, USA}
\author[0000-0001-8537-9299]{P.~J.~Boyle}
  \affiliation{Department of Physics, McGill University, 3600 rue University, Montr\'eal, QC H3A 2T8, Canada}
  \affiliation{Trottier Space Institute, McGill University, 3550 rue University, Montr\'eal, QC H3A 2A7, Canada}
\author[0000-0002-1800-8233]{Charanjot Brar}
  \affiliation{Department of Physics, McGill University, 3600 rue University, Montr\'eal, QC H3A 2T8, Canada}
  \affiliation{Trottier Space Institute, McGill University, 3550 rue University, Montr\'eal, QC H3A 2A7, Canada}
\author[0000-0002-2349-3341]{Daniela Breitman}
  \affiliation{Scuola Normale Superiore, Piazza dei Cavalieri 7, I-56126 Pisa, Italy}
\author[0000-0003-2047-5276]{Tomas Cassanelli}
  \affiliation{Department of Electrical Engineering, Universidad de Chile, Av. Tupper 2007, Santiago 8370451, Chile}
\author[0000-0002-3426-7606]{Pragya Chawla}
  \affiliation{Anton Pannekoek Institute for Astronomy, University of Amsterdam, Science Park 904, 1098 XH Amsterdam, The Netherlands}
\author[0000-0001-6422-8125]{Amanda M.~Cook}
  \affiliation{David A.~Dunlap Department of Astronomy \& Astrophysics, University of Toronto, 50 St.~George Street, Toronto, ON M5S 3H4, Canada}
  \affiliation{Dunlap Institute for Astronomy \& Astrophysics, University of Toronto, 50 St.~George Street, Toronto, ON M5S 3H4, Canada}
\author[0000-0002-8376-1563]{Alice P.~Curtin}
  \affiliation{Department of Physics, McGill University, 3600 rue University, Montr\'eal, QC H3A 2T8, Canada}
  \affiliation{Trottier Space Institute, McGill University, 3550 rue University, Montr\'eal, QC H3A 2A7, Canada}
\author[0000-0001-7166-6422]{Matt Dobbs}
  \affiliation{Department of Physics, McGill University, 3600 rue University, Montr\'eal, QC H3A 2T8, Canada}
  \affiliation{Trottier Space Institute, McGill University, 3550 rue University, Montr\'eal, QC H3A 2A7, Canada}
\author[0000-0003-4098-5222]{Fengqiu Adam Dong}
  \affiliation{Department of Physics and Astronomy, University of British Columbia, 6224 Agricultural Road, Vancouver, BC V6T 1Z1 Canada}
\author[0000-0003-3734-8177]{Gwendolyn Eadie}
  \affiliation{David A.~Dunlap Department of Astronomy \& Astrophysics, University of Toronto, 50 St.~George Street, Toronto, ON M5S 3H4, Canada}
  \affiliation{Department of Statistical Sciences}
  \affiliation{University of Toronto}
\author[0000-0001-8384-5049]{Emmanuel Fonseca}
  \affiliation{Department of Physics and Astronomy, West Virginia University, P.O. Box 6315, Morgantown, WV 26506, USA }
  \affiliation{Center for Gravitational Waves and Cosmology, West Virginia University, Chestnut Ridge Research Building, Morgantown, WV 26505, USA}
\author[0000-0002-3382-9558]{B.~M.~Gaensler}
  \affiliation{Dunlap Institute for Astronomy \& Astrophysics, University of Toronto, 50 St.~George Street, Toronto, ON M5S 3H4, Canada}
  \affiliation{David A.~Dunlap Department of Astronomy \& Astrophysics, University of Toronto, 50 St.~George Street, Toronto, ON M5S 3H4, Canada}
  \affiliation{Present address: Division of Physical and Biological Sciences, University of California Santa Cruz, 1156 High Street, Santa Cruz, CA 95064, USA}
\author[0000-0001-5553-9167]{Utkarsh Giri}
  \affiliation{Department of Physics, University of Wisconsin-Madison, 1150 University Ave, Madison, WI 53706, USA}
\author[0000-0002-3654-4662]{Antonio Herrera-Martin}
  \affiliation{David A.~Dunlap Department of Astronomy \& Astrophysics, University of Toronto, 50 St.~George Street, Toronto, ON M5S 3H4, Canada}
\author[0009-0002-1199-8876]{Hans Hopkins}
  \affiliation{Perimeter Institute for Theoretical Physics, 31 Caroline Street N, Waterloo, ON N25 2YL, Canada}
\author[0000-0003-2405-2967]{Adaeze L.~Ibik}
  \affiliation{David A.~Dunlap Department of Astronomy \& Astrophysics, University of Toronto, 50 St.~George Street, Toronto, ON M5S 3H4, Canada}
  \affiliation{Dunlap Institute for Astronomy \& Astrophysics, University of Toronto, 50 St.~George Street, Toronto, ON M5S 3H4, Canada}
\author[0000-0003-3457-4670]{Ronniy C.~Joseph}
  \affiliation{Department of Physics, McGill University, 3600 rue University, Montr\'eal, QC H3A 2T8, Canada}
  \affiliation{Trottier Space Institute, McGill University, 3550 rue University, Montr\'eal, QC H3A 2A7, Canada}
\author[0000-0003-4810-7803]{J.~F.~Kaczmarek}
  \affiliation{CSIRO Space \& Astronomy, Parkes Observatory, P.O. Box 276, Parkes NSW 2870, Australia}
\author[0000-0003-2739-5869]{Zarif Kader}
  \affiliation{Department of Physics, McGill University, 3600 rue University, Montr\'eal, QC H3A 2T8, Canada}
  \affiliation{Trottier Space Institute, McGill University, 3550 rue University, Montr\'eal, QC H3A 2A7, Canada}
\author[0000-0001-9345-0307]{Victoria M.~Kaspi}
  \affiliation{Department of Physics, McGill University, 3600 rue University, Montr\'eal, QC H3A 2T8, Canada}
  \affiliation{Trottier Space Institute, McGill University, 3550 rue University, Montr\'eal, QC H3A 2A7, Canada}
\author[0000-0003-2116-3573]{Adam E.~Lanman}
  \affiliation{MIT Kavli Institute for Astrophysics and Space Research, Massachusetts Institute of Technology, 77 Massachusetts Ave, Cambridge, MA 02139, USA}
  \affiliation{Department of Physics, Massachusetts Institute of Technology, 77 Massachusetts Ave, Cambridge, MA 02139, USA}
\author[0000-0002-5857-4264]{Mattias Lazda}
  \affiliation{David A.~Dunlap Department of Astronomy \& Astrophysics, University of Toronto, 50 St.~George Street, Toronto, ON M5S 3H4, Canada}
  \affiliation{Dunlap Institute for Astronomy \& Astrophysics, University of Toronto, 50 St.~George Street, Toronto, ON M5S 3H4, Canada}
\author[0000-0002-4209-7408]{Calvin Leung}
  \affiliation{Department of Astronomy, University of California Berkeley, Berkeley, CA 94720, USA}
  \affiliation{NHFP Einstein Fellow}
\author[0009-0000-9400-8609]{Siqi Liu}
  \affiliation{Department of Physics, McGill University, 3600 rue University, Montr\'eal, QC H3A 2T8, Canada}
  \affiliation{Trottier Space Institute, McGill University, 3550 rue University, Montr\'eal, QC H3A 2A7, Canada}
\author[0000-0002-4279-6946]{Kiyoshi W.  Masui}
  \affiliation{MIT Kavli Institute for Astrophysics and Space Research, Massachusetts Institute of Technology, 77 Massachusetts Ave, Cambridge, MA 02139, USA}
  \affiliation{Department of Physics, Massachusetts Institute of Technology, 77 Massachusetts Ave, Cambridge, MA 02139, USA}
\author[0000-0001-7348-6900]{Ryan Mckinven}
  \affiliation{Department of Physics, McGill University, 3600 rue University, Montr\'eal, QC H3A 2T8, Canada}
  \affiliation{Trottier Space Institute, McGill University, 3550 rue University, Montr\'eal, QC H3A 2A7, Canada}
\author[0000-0002-0772-9326]{Juan Mena-Parra}
  \affiliation{Dunlap Institute for Astronomy \& Astrophysics, University of Toronto, 50 St.~George Street, Toronto, ON M5S 3H4, Canada}
  \affiliation{David A.~Dunlap Department of Astronomy \& Astrophysics, University of Toronto, 50 St.~George Street, Toronto, ON M5S 3H4, Canada}
\author[0000-0003-2095-0380]{Marcus Merryfield}
  \affiliation{Department of Physics, McGill University, 3600 rue University, Montr\'eal, QC H3A 2T8, Canada}
  \affiliation{Trottier Space Institute, McGill University, 3550 rue University, Montr\'eal, QC H3A 2A7, Canada}
\author[0000-0002-2551-7554]{Daniele Michilli}
  \affiliation{MIT Kavli Institute for Astrophysics and Space Research, Massachusetts Institute of Technology, 77 Massachusetts Ave, Cambridge, MA 02139, USA}
  \affiliation{Department of Physics, Massachusetts Institute of Technology, 77 Massachusetts Ave, Cambridge, MA 02139, USA}
\author[0000-0002-3616-5160]{Cherry Ng}
  \affiliation{Laboratoire de Physique et Chimie de l'Environnement et de l'Espace - Universit\'e d'Orl\'eans/CNRS, 45071, Orl\'eans Cedex 02, France}
\author[0000-0003-0510-0740]{Kenzie Nimmo}
  \affiliation{MIT Kavli Institute for Astrophysics and Space Research, Massachusetts Institute of Technology, 77 Massachusetts Ave, Cambridge, MA 02139, USA}
\author[0000-0002-5254-243X]{Gavin Noble}
  \affiliation{David A.~Dunlap Department of Astronomy \& Astrophysics, University of Toronto, 50 St.~George Street, Toronto, ON M5S 3H4, Canada}
  \affiliation{Dunlap Institute for Astronomy \& Astrophysics, University of Toronto, 50 St.~George Street, Toronto, ON M5S 3H4, Canada}
\author[0000-0002-8897-1973]{Ayush Pandhi}
  \affiliation{David A.~Dunlap Department of Astronomy \& Astrophysics, University of Toronto, 50 St.~George Street, Toronto, ON M5S 3H4, Canada}
  \affiliation{Dunlap Institute for Astronomy \& Astrophysics, University of Toronto, 50 St.~George Street, Toronto, ON M5S 3H4, Canada}
\author[0000-0003-3367-1073]{Chitrang Patel}
  \affiliation{Department of Physics, McGill University, 3600 rue University, Montr\'eal, QC H3A 2T8, Canada}
  \affiliation{Trottier Space Institute, McGill University, 3550 rue University, Montr\'eal, QC H3A 2A7, Canada}
\author[0000-0002-8912-0732]{Aaron B.~Pearlman}
  \affiliation{Department of Physics, McGill University, 3600 rue University, Montr\'eal, QC H3A 2T8, Canada}
  \affiliation{Trottier Space Institute, McGill University, 3550 rue University, Montr\'eal, QC H3A 2A7, Canada}
  \affiliation{Banting Fellow}
  \affiliation{McGill Space Institute Fellow}
  \affiliation{FRQNT Postdoctoral Fellow}
\author[0000-0003-2155-9578 ]{Ue-Li Pen}
  \affiliation{Canadian Institute for Theoretical Astrophysics, 60 St.~George Street, Toronto, ON M5S 3H8, Canada}
  \affiliation{Department of Physics, University of Toronto, 60 St.~George Street, Toronto, ON M5S 1A7, Canada}
\author[0000-0002-9822-8008]{Emily Petroff}
  \affiliation{Perimeter Institute for Theoretical Physics, 31 Caroline Street N, Waterloo, ON N25 2YL, Canada}
\author[0000-0002-4795-697X]{Ziggy Pleunis}
  \affiliation{Dunlap Institute for Astronomy \& Astrophysics, University of Toronto, 50 St.~George Street, Toronto, ON M5S 3H4, Canada}
\author[0000-0001-7694-6650]{Masoud Rafiei-Ravandi}
  \affiliation{Department of Physics, McGill University, 3600 rue University, Montr\'eal, QC H3A 2T8, Canada}
  \affiliation{Trottier Space Institute, McGill University, 3550 rue University, Montr\'eal, QC H3A 2A7, Canada}
\author[0000-0003-1842-6096]{Mubdi Rahman}
  \affiliation{Sidrat Research, 124 Merton Street, Suite 507, Toronto, ON M4S 2Z2, Canada}
\author[0000-0001-5799-9714]{Scott M.~Ransom}
  \affiliation{National Radio Astronomy Observatory, 520 Edgemont Rd, Charlottesville, VA 22903, USA}
\author[0000-0003-3154-3676]{Ketan R.~Sand}
  \affiliation{Department of Physics, McGill University, 3600 rue University, Montr\'eal, QC H3A 2T8, Canada}
  \affiliation{Trottier Space Institute, McGill University, 3550 rue University, Montr\'eal, QC H3A 2A7, Canada}
\author[0000-0002-7374-7119]{Paul Scholz}
  \affiliation{Department of Physics and Astronomy, York University, 4700 Keele Street, Toronto, ON MJ3 1P3, Canada}
  \affiliation{Dunlap Institute for Astronomy \& Astrophysics, University of Toronto, 50 St.~George Street, Toronto, ON M5S 3H4, Canada}
\author[0000-0002-4823-1946]{Vishwangi Shah}
  \affiliation{Department of Physics, McGill University, 3600 rue University, Montr\'eal, QC H3A 2T8, Canada}
  \affiliation{Trottier Space Institute, McGill University, 3550 rue University, Montr\'eal, QC H3A 2A7, Canada}
\author[0000-0002-6823-2073]{Kaitlyn Shin}
  \affiliation{MIT Kavli Institute for Astrophysics and Space Research, Massachusetts Institute of Technology, 77 Massachusetts Ave, Cambridge, MA 02139, USA}
  \affiliation{Department of Physics, Massachusetts Institute of Technology, 77 Massachusetts Ave, Cambridge, MA 02139, USA}
\author[0009-0000-6294-5315]{Yuliya Shpunarska}
  \affiliation{Department of Physics, McGill University, 3600 rue University, Montr\'eal, QC H3A 2T8, Canada}
\author[0000-0003-2631-6217]{Seth R.~Siegel}
  \affiliation{Perimeter Institute for Theoretical Physics, 31 Caroline Street N, Waterloo, ON N25 2YL, Canada}
  \affiliation{Department of Physics, McGill University, 3600 rue University, Montr\'eal, QC H3A 2T8, Canada}
  \affiliation{Trottier Space Institute, McGill University, 3550 rue University, Montr\'eal, QC H3A 2A7, Canada}
\author[0000-0002-2088-3125]{Kendrick Smith}
  \affiliation{Perimeter Institute for Theoretical Physics, 31 Caroline Street N, Waterloo, ON N25 2YL, Canada}
\author[0000-0001-9784-8670]{Ingrid Stairs}
  \affiliation{Department of Physics and Astronomy, University of British Columbia, 6224 Agricultural Road, Vancouver, BC V6T 1Z1 Canada}
\author[0000-0002-9761-4353]{David C.~Stenning}
  \affiliation{Department of Statistics \& Actuarial Science, Simon Fraser University, 8888 University Dr, Burnaby, BC V5A 1S6, Canada}
\author[0000-0003-4535-9378]{Keith Vanderlinde}
  \affiliation{Dunlap Institute for Astronomy \& Astrophysics, University of Toronto, 50 St.~George Street, Toronto, ON M5S 3H4, Canada}
  \affiliation{David A.~Dunlap Department of Astronomy \& Astrophysics, University of Toronto, 50 St.~George Street, Toronto, ON M5S 3H4, Canada}
\author[0000-0002-1491-3738]{Haochen Wang}
  \affiliation{MIT Kavli Institute for Astrophysics and Space Research, Massachusetts Institute of Technology, 77 Massachusetts Ave, Cambridge, MA 02139, USA}
  \affiliation{Department of Physics, Massachusetts Institute of Technology, 77 Massachusetts Ave, Cambridge, MA 02139, USA}
\author[0000-0003-1771-2218]{Henry White}
  \affiliation{Sidrat Research, 124 Merton Street, Suite 507, Toronto, ON M4S 2Z2, Canada}
\author[0000-0001-7314-9496]{Dallas Wulf}
  \affiliation{Department of Physics, McGill University, 3600 rue University, Montr\'eal, QC H3A 2T8, Canada}
  \affiliation{Trottier Space Institute, McGill University, 3550 rue University, Montr\'eal, QC H3A 2A7, Canada}
\newcommand{\allacks}{
B.C.A.\ is supported by an FRQNT Doctoral Research Award.
M.B.\ is a McWilliams fellow and an International Astronomical Union Gruber fellow. M.B. also receives support from the McWilliams seed grant.
A.M.C.\ is funded by an NSERC Doctoral Postgraduate Scholarship. 
A.P.C.\ is a Vanier Canada Graduate Scholar.
M.D.\ is supported by a CRC Chair, NSERC Discovery Grant, CIFAR, and by the FRQNT Centre de Recherche en Astrophysique du Qu\'ebec (CRAQ).
F.A.D.\ is supported by the UBC Four Year Fellowship.
G.M.E.\ acknowledges support from a Collaborative Research Team Grant from the Canadian Statistical Sciences Institute, and the support of the Natural Sciences and Engineering Research Council of Canada (NSERC) through grant RGPIN-2020-04554
B.M.G.\ acknowledges the support of the Natural Sciences and Engineering Research Council of Canada (NSERC) through grant RGPIN-2022-03163, and of the Canada Research Chairs program.
A.H.M.\ is supported by a CANSSI Collaborative Research Team Grant.
V.M.K.\ holds the Lorne Trottier Chair in Astrophysics \& Cosmology, a Distinguished James McGill Professorship, and receives support from an NSERC Discovery grant (RGPIN 228738-13), from an R. Howard Webster Foundation Fellowship from CIFAR, and from the FRQNT CRAQ.
C.L.\ is supported by NASA through the NASA Hubble Fellowship grant HST-HF2-51536.001-A awarded by the Space Telescope Science Institute, which is operated by the Association of Universities for Research in Astronomy, Inc., under NASA contract NAS5-26555.
K.W.M. holds the Adam J. Burgasser Chair in Astrophysics and is supported by NSF grants (2008031, 2018490).
M.M.\ is supported by an NSERC PGS-D award.
K.N.\ is an MIT Kavli Fellow.
A.P.\ is funded by the NSERC Canada Graduate Scholarships -- Doctoral program.
A.B.P.\ is a Banting Fellow, a McGill Space Institute~(MSI) Fellow, and a Fonds de Recherche du Quebec -- Nature et Technologies~(FRQNT) postdoctoral fellow.
Z.P.\ is a Dunlap Fellow.
S.M.R.\ is a CIFAR Fellow and is supported by the NSF Physics Frontiers Center award 2020265. 
K.R.S.\ is supported by FRQNT doctoral research award.
K.S.\ is supported by the NSF Graduate Research Fellowship Program.
D.C.S.\ is supported by an NSERC Discovery Grant (RGPIN-2021-03985) and by a Canadian Statistical Sciences Institute (CANSSI) Collaborative Research Team Grant.
}